\documentclass[sigconf]{acmart}

\pdfoutput=1

\usepackage[lined, noend, linesnumbered]{algorithm2e}
\SetKwBlock{Init}{}{}

\SetCommentSty{mycommfont}

\usepackage{tikz}
\usepackage{xcolor}
\usepackage{caption}
\usepackage{subcaption}
\usepackage{cleveref}
\usepackage{mathrsfs}
\usepackage{amsmath}
\providecommand{\coloneqq}{\mathrel{\mathop:}=}
\usepackage{amsthm}
\usepackage{thmtools}

\setcopyright{none}
\renewcommand\footnotetextcopyrightpermission[1]{}
\acmConference{}{}{}

\begin{document}

\newcommand{\name}{Ditto}

\title{\name: Generalized Reconfigurable Linearizable Reads}

\author{Aleksey Panas}
\email{aleksey.panas@alumni.utoronto.ca}
\orcid{0009-0009-5824-2081}
\affiliation{%
  \institution{University of Toronto}
  \city{Toronto}
  \state{Ontario}
  \country{Canada}
}

\renewcommand{\shortauthors}{Aleksey Panas}

\begin{abstract}
    
Linearizability gives developers the illusion that operations on a distributed datastore execute sequentially on a single machine. Providing it is expensive, so numerous specialized reads algorithms have been proposed to make reads significantly faster without sacrificing linearizability, which matters because most workloads are read-dominant by orders of magnitude. Yet, none of these algorithms is the best choice for every deployment. We show why with a framework that classifies these algorithms into a generalized space together with a mathematical analysis of their latencies as a function of this generalization. The analysis exposes two fundamental tradeoffs, one between read and write latency and one between read and write tolerance to network variance, which together leave the best algorithm dependent on network and workload conditions that vary over time. It also exposes a gap: there is a way to minimize the delay a read incurs when it overlaps a concurrent write, at no cost to writes, yet existing algorithms can apply it only for a subset of the tradeoff space, forcing any deployment that needs the full space to settle for a strictly worse solution. We close this gap with a new algorithm, Pairwise Quorums, and build a second, Ditto, that reconfigures across the space at runtime as conditions change. Together they subsume every existing algorithm under our model: against each, the combination is either strictly better, or reconfigures to match it where a genuine tradeoff applies.
\end{abstract}

\maketitle
\pagestyle{plain}

\section{Introduction}
\label{sec:introduction}

Linearizable datastores such as Spanner \cite{corbett2013spanner}, CockroachDB \cite{taft2020cockroachdb}, and YugabyteDB are now widely deployed in industry. Linearizability \cite{herlihy1990linearizability} is desirable because it gives applications the illusion of a single, sequential machine, letting developers build geo-distributed, fault-tolerant systems without reasoning about concurrency. To provide it, these systems run a state machine replication (SMR) protocol such as MultiPaxos \cite{van2015paxos} or Raft \cite{ongaro2014search}, which exposes an interface for executing read and write operations on replicated application state.

Unfortunately, executing an operation in SMR is expensive, requiring rounds of communication to reach consensus, and a single elected leader to sequence operations. To escape this cost, developers often fall back to weaker models such as eventual consistency, trading strong guarantees for performance.

\begin{figure}[htbp]
    \centering
    \includegraphics[width=0.48\textwidth]{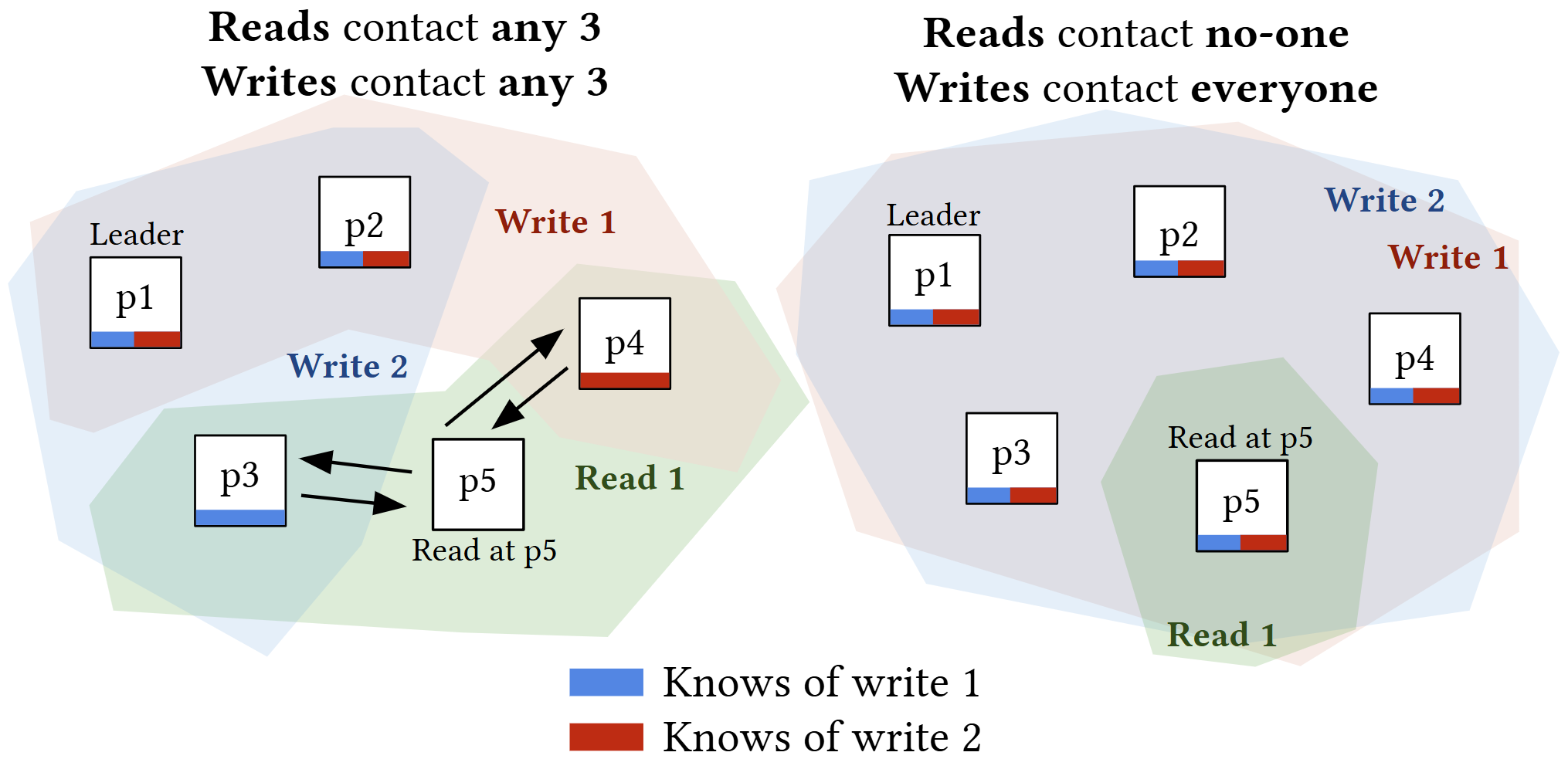}
    \caption{Decreasing the size of \textit{read quorums} increases size of \textit{write quorums}}
    \label{fig:n_latency_tradeoff}
\end{figure}

However, the non-mutating property of read operations has led to the development of numerous \textit{specialized reads algorithms} which significantly improve read performance without sacrificing linearizability. These algorithms are of particular importance as most workloads are read-dominant. For example, Google reported that in their ads workload, read operations outnumber write operations by three orders of magnitude \cite{corbett2013spanner}. Today's existing specialized reads algorithms include Leader Reads \cite{chandra2007paxos}, PQL \cite{moraru2014paxos}, PQR \cite{charapko2019linearizable}, EdgePQR \cite{guarnieri2023linearizable}, CHT \cite{chandra2016algorithm}, and Pairwise \cite{thiessen2025asymmetric}.

In this paper, we show that none of these reads algorithms is the best choice for every deployment by presenting a framework that decomposes each algorithm into a combination of two independent choices. The first is the read-write quorum system \cite{whittaker2021read}, dictating which processes a read contacts, and consequently which a write must contact so that no read misses a completed write. The second is how a read behaves when it executes concurrently with a write, which is called the \textit{stamping mechanism}. 

Taking the quorum system first, this choice encodes two fundamental tradeoffs. The first is read-write latency. To preserve linearizability, a read must observe every write that completed before it began, which requires contacting at least one process aware of each such write. Formally, every write quorum (the processes contacted for a write) must intersect every read quorum (the processes contacted for a read) \cite{whittaker2021read}. Shrinking read quorums to lower read latency therefore forces write quorums to grow, and vice versa. PQR \cite{charapko2019linearizable}, for example, lets any replica serve a read by contacting a simple majority making read latency high, but leaving writes unaffected since they already contact a majority under standard SMR. Pairwise \cite{thiessen2025asymmetric} sits at the opposite extreme, serving fully local reads at every replica at the cost of requiring writes to contact all processes (Figure \ref{fig:n_latency_tradeoff}).

The second tradeoff is tolerance to network variance, that is, variance in the latency of inter-process links. Consider a system running Pairwise when a process becomes fail-slow, exhibiting a prolonged latency increase \cite{gunawi2018fail}. Because the leader must contact every process on every write, write progress degrades until a costly recovery is triggered. PQR instead gives the leader a choice of which processes to contact to satisfy quorum intersection, tolerating such variance, but at the price of higher read latency.

As a consequence of these tradeoffs, the best quorum system configuration depends on workload and network conditions. A read-heavy system over good links wants small read quorums, while a write-heavy or flaky one wants large ones. Worse, a single system wants different settings at different times. It may be read-dominant by day, when users are active, and write-dominant by night, as new content lands, and its links may degrade and recover throughout. A system should be able to move along this space as conditions change, but no current algorithm concretely provides this.

The choice of stamping mechanism tells a different story. When a read occurs concurrently with an in-progress write, the process serving it determines whether to \textit{stamp} with that write, forcing the read to wait for that write to finish, or to stamp with an older write and return the read sooner. This decision is tied to correctness, since stamping incorrectly may cause a read to miss a completed write, violating linearizability. This period of waiting for the write the read incurs is called read blockage time \cite{bi2022parameterized}, and the stamping mechanism is what controls it. Here one option, delayed stamping, is best. It gives the lowest read blockage time at no cost to writes. However, existing reads algorithms that implement delayed stamping are only capable of quorum system configurations involving local readers, that is, read quorums of size one. A system that wants a quorum-based setting, say to tolerate network variance for writes, is forced onto a worse stamping mechanism and pays blockage time it should not have to.

Put together, the two choices explain why there is no clear winner. Each algorithm is beaten wherever its quorum configuration is wrong for the current network and workload conditions, or its stamping mechanism is not the best one available. Winning everywhere means getting both choices right at once. Today, this is not possible.

To address the issue, we contribute two novel algorithms. The first, Pairwise Quorums, extends delayed stamping to support any quorum system configuration, forever removing the need to sacrifice read blockage time in exchange for more extensive quorum system support. The second algorithm, Ditto, provides a black box at-runtime quorum system reconfiguration solution, allowing the system to dynamically adapt to the current deployment conditions and always maintain the optimal configuration. The collective combination of Ditto and Pairwise Quorums is an algorithm which subsumes all existing algorithms under our model. This paper extends our earlier preprint ~\cite{thiessen2024towards}, which proposed switching among read-write quorum systems at runtime.

Of course, Ditto's reconfiguration ability on its own does not answer the question of when reconfiguration should happen. In order to provide clarity within the quorum system configuration space, we provide a mathematical analysis of \textit{write latency}, \textit{best case read latency}, \textit{worst case read latency}, and \textit{network variance} as a function of the quorum configuration, motivating our two algorithms and serving as a precise tool for the developer to determine the performance of a system a priori under given conditions and configuration.

The paper proceeds as follows. Section \ref{sec:background} covers background and introduces every existing linearizable reads algorithm. Section \ref{sec:categorization} places them in our framework, analyzes them, and identifies the gap in stamping mechanisms. Section \ref{sec:extending} presents and analyzes Pairwise-Quorums. Section \ref{sec:reconfig} presents Ditto and reconfiguration. Section \ref{sec:experiments} validates the analysis experimentally and shows the performance gains of Pairwise-Quorums and Ditto. Section \ref{sec:related} presents related work and section \ref{sec:conclusion} concludes and discusses future work.

\section{Background}
\label{sec:background}

In this section, we discuss the system model, state machine replication, existing algorithms for linearizable reads, and their relationship to read-write quorum systems.

\subsection{Model}
\label{subsec:model}

We assume an asynchronous system of $n$ processes, of which at most $f < \frac{n}{2}$ may crash. Messages may be arbitrarily delayed, reordered, or lost. We additionally require that processes can grant \textit{leases} to tolerate failures without sacrificing liveness~\cite{gray1989leases}. A lease must be granted \textit{correctly}, that is, the granter must perceive the lease as expired only after the holder does, in real time. This requires assuming either synchronized clocks with bounded skew~\cite{gray1989leases} or hardware clocks with bounded drift~\cite{liskov1991practical}.

\subsection{State Machine Replication}
\label{sec:smr}

State machine replication (SMR) is the de facto standard for linearizable, fault-tolerant replication of client operations across a set of processes. Linearizability totally orders operations in a way that respects their real-time order: if $o_1$ completes before $o_2$ begins, then $o_1$ is ordered before $o_2$~\cite{herlihy1990linearizability}.

Each process holds a \textit{replica} of the application state, and replicas are deterministic in that applying the same sequence of operations to any replica yields the same state~\cite{schneider1990implementing}. Clients submit operations to a process, which runs the read or write procedure and returns the result. To order writes, SMR relies on a distinguished \textit{leader} that proceeds in two phases. In the \textit{prepare} phase, the leader proposes assigning a write $w$ to index $i$ of the \textit{log}; the phase completes, and $i$ is \textit{decided}, once a majority accept. The leader then \textit{commits} $w$, and a process applies $w$ to its replica once it has applied every write up to index $i$.

A read can be served naively as a no-op write. Since SMR writes are read-modify-writes (RMW), this returns the latest state without changing it, but at the cost of a full write. Of course, this is inefficient and the \textit{specialized reads algorithms} we discuss next avoid this. 

\subsection{Local Read Algorithms}

\begin{figure}[htbp]
    \centering
    \includegraphics[width=0.42\textwidth]{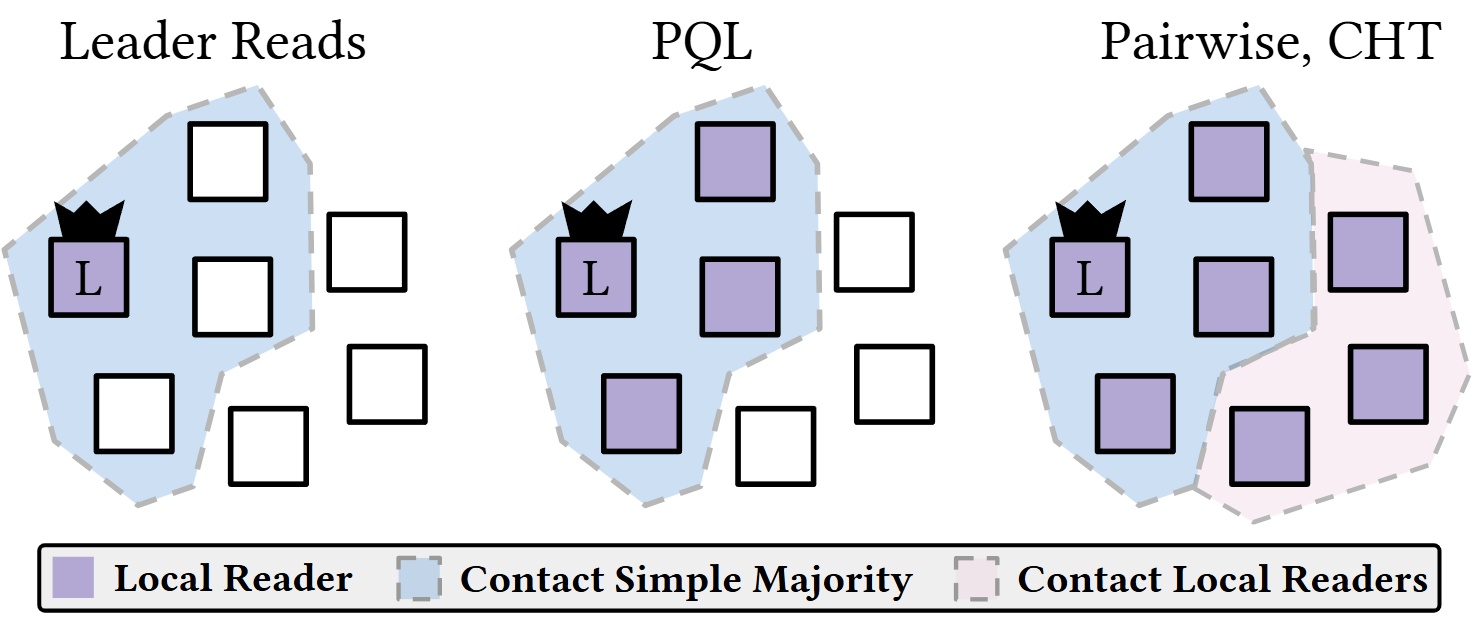}
    \caption{A write contacts any simple majority AND all local readers. The two conditions may overlap, as seen in all three cases above.}
    \label{fig:n_local_rq_comparison}
\end{figure}

\textit{Local Reads} algorithms, including PQL~\cite{moraru2014paxos}, CHT~\cite{chandra2016algorithm}, Leader Reads~\cite{chandra2007paxos}, and Pairwise~\cite{thiessen2025asymmetric}, designate a subset of processes as \textit{local readers}. A local reader serves reads without contacting any other process; every other process forwards its reads to the closest local reader. To ensure reads observe every completed write, the leader must then contact all local readers on each write, in addition to a majority. Figure \ref{fig:n_local_rq_comparison} shows how each algorithm places its local readers and the resulting \textit{quorums}.

Since writes wait on specific local readers, the failure of one would stall write progress. To prevent this, a local reader serves reads only while holding a \textit{read lease}~\cite{moraru2014paxos, gray1989leases}. If a local reader fails, the leader invalidates its lease by contacting a majority, or waits for it to expire, and proceeds without it; a recovered reader can later re-establish its lease safely.

A local reader $p$ serves a read by assigning it an index $i$ of some RMW, called \textit{stamping} with $i$. Correctness requires $i$ to be at least the index of the latest completed RMW. The stamped read may then be returned by any process that has committed every write up to and including $i$, not necessarily $p$ itself. This makes reads and RMWs linearizable by completion order: RMW $i$ completes, then all reads stamped with $i$, then RMW $i+1$, and so on.

Existing local read algorithms use one of three \textit{stamping mechanisms} to assign $i$: Invalidation, Eager Stamping, and Delayed Stamping. We compare them using \textit{stop} and \textit{go} events~\cite{thiessen2025asymmetric}. A \textit{stop event} for index $i$ is the point on $p$'s clock after which $p$ must stamp reads with indices $\geq i$; intuitively, $p$ has learned of a write $i$ whose completion is uncertain and must conservatively include it. The \textit{go event} for $i$ is the earliest point at which $p$ may return state through index $i$, once $i$ is committed. For correctness, the stop event for $i$ on every local reader must precede write $i$'s completion at the leader and the go event must follow it. Reads stamped with $i$ block during the interval between the two events; we call its length the \textit{read blockage time}, and minimizing it is desirable.

\begin{figure}[htbp]
    \centering
    \includegraphics[width=0.49\textwidth]{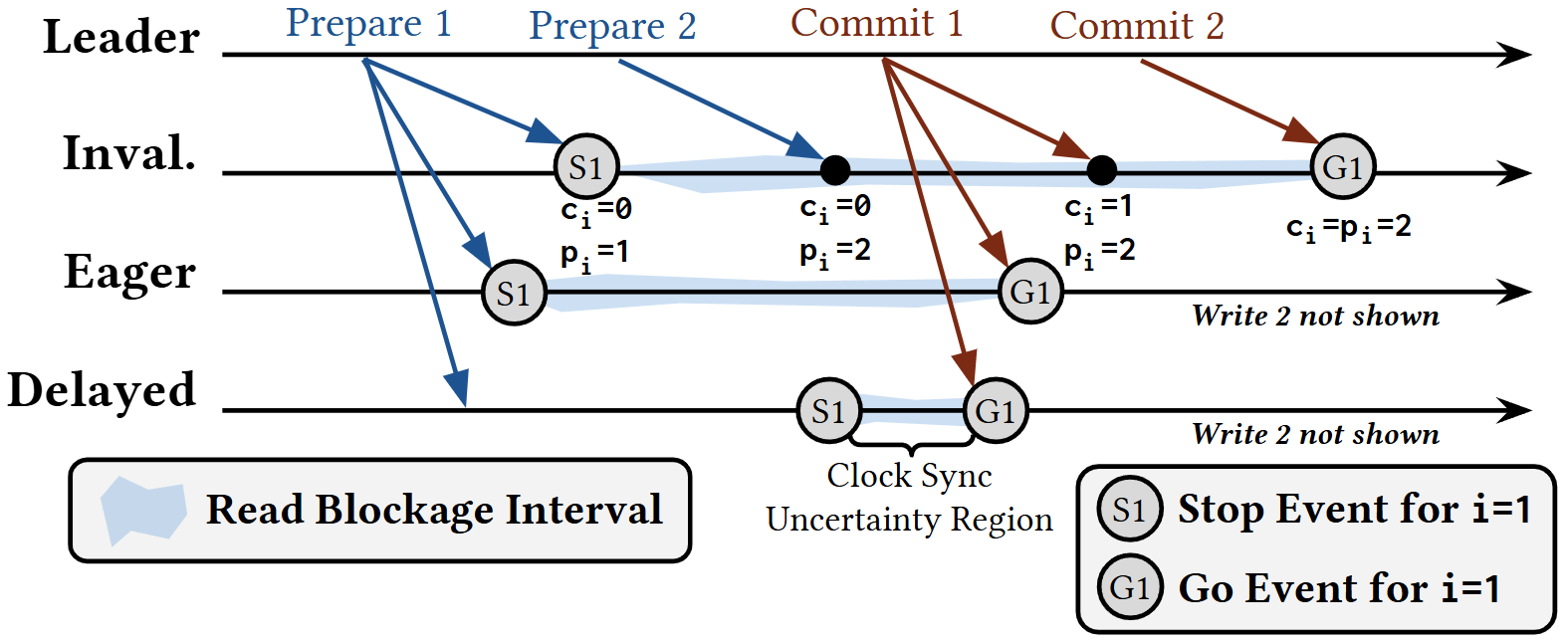}
    \caption{In the worst case, Invalidation blocks for multiple possibly infinite writes, Eager blocks for one write, and Delayed often blocks for less than one write}
    \label{fig:n_stamping_comparison}
\end{figure}

We now describe the three mechanisms, summarized in Figure \ref{fig:n_stamping_comparison}.

\subsubsection{Invalidation} Used by PQL~\cite{moraru2014paxos}, Invalidation places the stop event for $i$ when $p$ receives the prepare for $i$, and the go event when $p$'s latest committed index equals its latest received prepare index. If writes are proposed continuously, the latest prepare index keeps outpacing the latest committed index, so the go event may never fire and reads can block indefinitely.

\subsubsection{Eager Stamping} Used by CHT~\cite{chandra2016algorithm}, Eager Stamping keeps Invalidation's stop event but sets the go event when $p$ receives the commit for $i$. The go event is now fixed, so read blockage is finite. In the worst case a read blocks for the full duration of the write.

\subsubsection{Delayed Stamping} Used by Pairwise~\cite{thiessen2025asymmetric} and Leader \\ Reads~\cite{chandra2007paxos}, Delayed Stamping has the leader assign each write $i$ a future \textit{promise time} $t+\alpha$ and promise not to commit $i$ before it, attaching this promise to prepare messages. A local reader then schedules its stop event for the latest time it can be sure $t+\alpha$ has not yet elapsed on the leader's clock, and its go event for the earliest time it can be sure it has. Under perfect clock synchronization the two coincide at $t+\alpha$, giving zero read blockage. Otherwise, only the leader achieves zero, and uncertainty grows with distance from it. We detail this mechanism for Pairwise in Section~\ref{subsec:pairwise}.

\subsection{Read-Write Quorum Systems}

Guaranteeing that reads observe all completed writes requires read-write quorum intersection. A \textit{read-write quorum system} is a set of read quorums and write quorums in which every write quorum intersects every read quorum~\cite{whittaker2021read}. A write completes once the leader hears from a full write quorum, and a read completes once it contacts a full read quorum. In practice it suffices to define the read quorums alone. A valid write quorum is then any majority containing at least one process from every read quorum. For example, with processes $\{A,B,C,D,E\}$, leader $A$, and read quorums $\{\{A,B\},\{D,E\}\}$, the set $\{B,C,D\}$ is a valid write quorum. Then, Local Reads algorithms are those with read quorums of size one.

\subsection{Quorum Read Algorithms}

\textit{Quorum Reads} algorithms are ones that use read quorums larger than one and include PQR~\cite{charapko2019linearizable} and EdgePQR~\cite{guarnieri2023linearizable}. In PQR, a process $p$ serves a read by contacting any majority to retrieve their latest prepare indexes, stamping with the maximum index $i_{\max}$, and returning state through $i_{\max}$ once committed on $p$. PQR's read quorums are thus any majority, and it uses a quorum-generalized variant of Eager Stamping. EdgePQR instead takes an explicitly defined read quorum set $Q^R$. $p$ contacts any $Q \in Q^R$ and follows PQR's procedure, with write quorums adjusted to intersect every read quorum. Like local reads, EdgePQR's smaller-than-majority quorums reduce fault tolerance, which is addressed with a \textit{read lease} mechanism extended to arbitrary-sized quorums.

\subsection{Pairwise}
\label{subsec:pairwise}

\begin{figure}[htbp]
    \centering
    \includegraphics[width=0.47\textwidth]{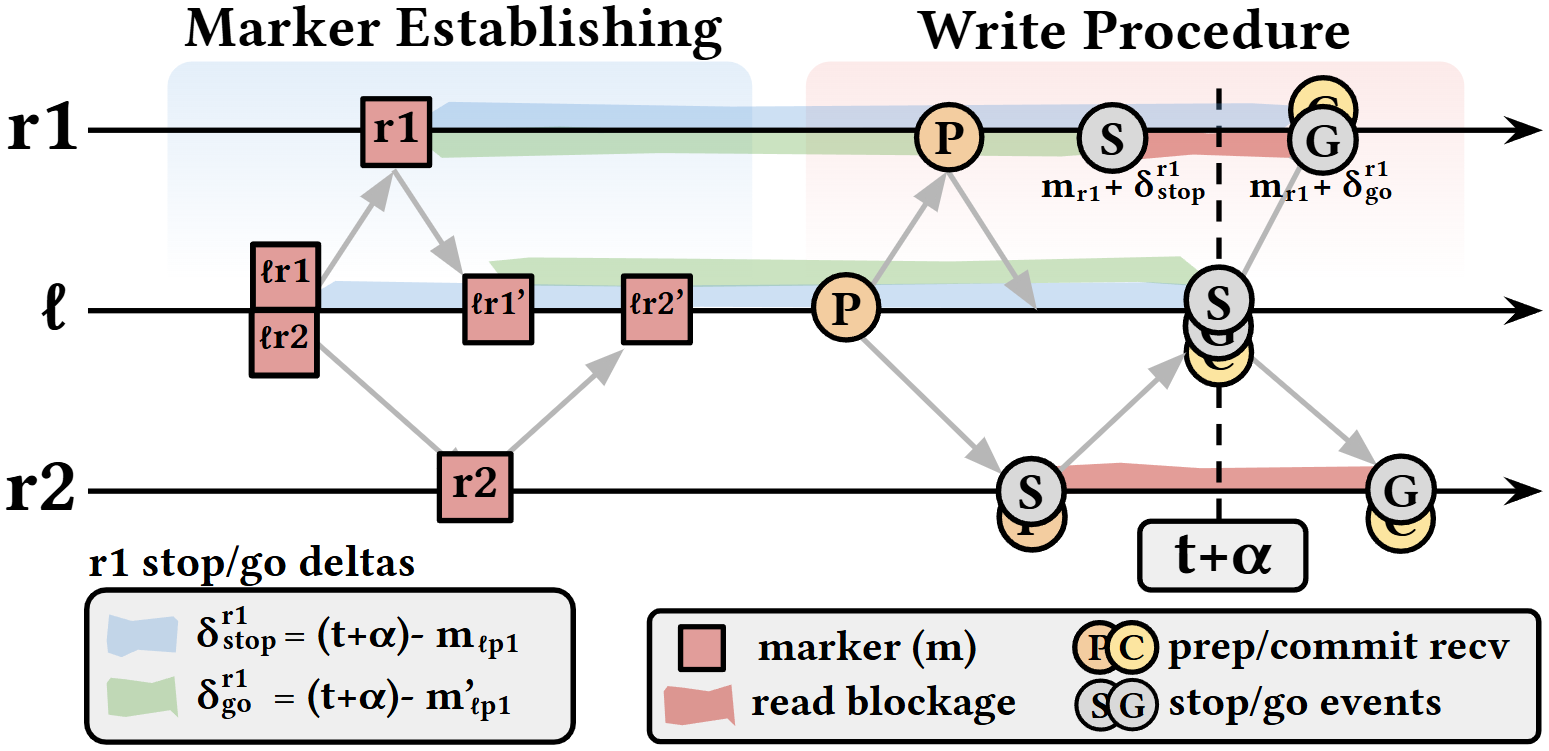}
    \caption{Leader $\ell$ uses marker mechanism to schedule stop and go events before and after promise time $t+\alpha$ on local reader $r1$ and $r2$. Read blockage time tends to $0$ with proximity to leader}
    \label{fig:n_pairwise}
\end{figure}

We now describe how Pairwise~\cite{thiessen2025asymmetric} achieves delayed stamping, assuming no clock drift for concision.

At startup, the leader $\ell$ establishes \textit{markers} with every local reader $p$: $\ell$ records a timestamp $m_{p\ell}$ and messages $p$, which records $m_p$ and replies, after which $\ell$ records $m'_{p\ell}$. To process write $i$, $\ell$ takes the current time $t$, and sets the promise time $t+\alpha$, promising not to commit before it. $\ell$ then attaches $\delta_{stop} = t+\alpha - m'_{p\ell}$ and $\delta_{go} = t+\alpha - m_{p\ell}$ to each prepare, and on receipt $p$ computes $stop_i = m_p + \delta_{stop}$ and $go_i = m_p + \delta_{go}$ as points on its own clock. The stop event for $i$ is then the later of $stop_i$ and the time $p$ receives the prepare for $i$, and the go event the later of $go_i$ and the time $p$ receives the commit for $i$. The markers guarantee that every stop event elapses on its reader before the promise time elapses at the leader, and every go event after, so reads correctly observe completed writes.

$\alpha$ does not affect correctness. Pairwise shows its ideal value is the RTT from $\ell$ to the furthest process $\ell$ must await on a write, so that $t+\alpha$ elapses exactly when the necessary replies arrive and the write incurs no extra latency. As Figure \ref{fig:n_pairwise} illustrates, the worst-case blockage time at a local reader $p$ is then the RTT from $\ell$ to $p$, namely $m'_{\ell p} - m_{\ell p}$.

\section{The Space of Linearizable Reads}
\label{sec:categorization}

\begin{figure}[htbp]
    \centering
    \includegraphics[width=0.4\textwidth]{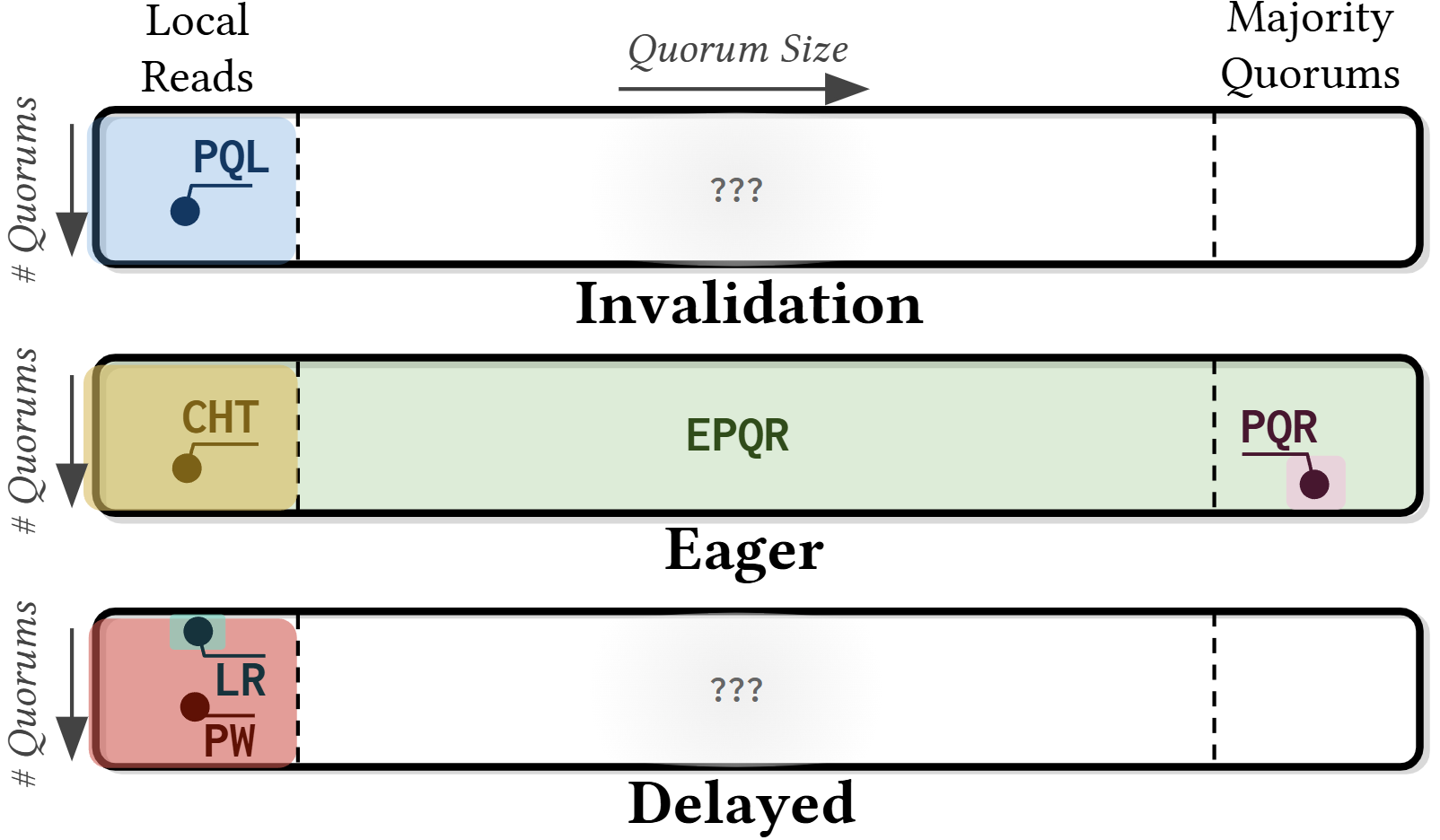}
    \caption{High level classification of existing algorithms shows gaps in the space of delayed stamping and invalidation}
    \label{fig:n_classification_table_pre}
\end{figure}

We now present our framework for classifying and analyzing all specialized read algorithms. As introduced in Section \ref{sec:introduction}, every algorithm is a combination of a stamping mechanism and a set of supported read-write quorum systems, summarized in Figure \ref{fig:n_classification_table_pre}. For example, PQR uses eager stamping with a single choice of setting read quorums to be every simple majority, while PQL uses invalidation with read quorums configurable to any group of local readers.

\subsubsection{General Write Procedure}
\label{subsubsec:general_write_alg}
In every algorithm, the write procedure sends prepares to all processes, waits for acknowledgments from a majority that intersects every read quorum, and broadcasts a commit.

\subsubsection{General Read Procedure}
\label{subsubsec:general_read_alg}
Conversely, the read procedure at process $p$ (1) broadcasts and waits for replies from a full read quorum, the closest one to $p$ determining the latency, (2) collects stop event information from the replies, (3) computes a stamp index $i$ as some aggregation over the replies, and (4) returns state through $i$ once $i$ is committed at $p$ and $p$'s go event for $i$ has elapsed. Step 3 is trivial for local reads, where each read quorum is a single reader that hands $p$ the index directly.

As previewed in the introduction and shown in Figure \ref{fig:n_classification_table_pre}, there are gaps in the quorum systems supported by invalidation and delayed stamping, confining them to local reads. Since delayed stamping dominates the other mechanisms in read blockage time, only its gap is worth filling. We also exclude read quorums larger than a majority since base SMR already prevents write quorums from shrinking below a majority, so enlarging read quorums past that point only worsens reads with no gain for writes. The analysis that follows derives key metrics and then traces their trends, confirming that delayed stamping is best and exposing the quorum system's two core tradeoff axes, which leave no single configuration optimal as the network and workload conditions change.

We define a network $N$ by the following:
\begin{enumerate}
    \item Let $\Pi$ be the set of processes in the network, and $n=|\Pi|$.
    \item Let $Q^R$ be the algorithm's set of read quorums, and $R=\bigcup_{Q\in Q^R} Q$ the set of all \textit{readers}.
    \item Let $\delta_{p_1p_2}$ be the one-way latency from $p_1 \in \Pi$ to $p_2 \in \Pi$.
    \item Let $\delta_{pSp} = \max_{s \in S}\{\delta_{ps} + \delta_{sp}\}$ be the time for $p$ to message a set $S$ and receive replies.
    \item For $p \in \Pi$, its \textit{closest} process or quorum is the one with smallest RTT by $\delta$, i.e. the first to reply if $p$ broadcasts to everyone.
    \item Let $\ell \in \Pi$ be the leader, and $M$ the \textit{closest} simple majority to $\ell$.
    \item If the algorithm is a local reads algorithm, $Q^R$ is a set of size-one read quorums; let $L$ be its set of local readers.
\end{enumerate}

As a function of $N$, we derive write latency ($\Delta_w(N)$), best case read latency ($BC(\Delta_r(N))$), and worst case read latency under write contention ($WC(\Delta_r(N))$). For this part of the analysis $\delta$ is fixed during an operation meaning the network is static; we treat network variance separately afterward.

We measure write latency from when $\ell$ broadcasts prepares to just before it broadcasts commits, and read latency from when a process $p$ receives a read request to when it is about to return the read state. The gap between best and worst case read latency is exactly the read blockage time. The best case has no concurrent write occurring and has zero blockage, while the worst case has writes proposed continuously, so a read waits the maximum possible time for a concurrent write to commit.

Throughout, we assume reads use a \textit{full broadcast} approach as mentioned in the general read procedure \ref{subsubsec:general_read_alg}. That is, $p$ contacts every process and proceeds once a read quorum has replied. The \textit{closest} quorum replies first and thus dictates the latency. We now give each metric's expression per stamping mechanism, each followed by a short justification.

\subsection{Eager Stamping Analysis}

We begin with Eager Stamping, which covers all quorum systems and so gives the most general expressions in terms of $Q^R$.

\subsubsection{Write Latency}
\begin{equation}
    \label{eq:eag-write-lat}
    \Delta_w(N)= \max(  \max_{Q \in Q^R}\{ \min_{q \in Q}( \delta_{\ell q} + \delta_{q \ell} ) \}, \delta_{\ell M \ell} )
\end{equation}

To commit a write, $\ell$ broadcasts prepares and waits for replies from (a) any majority and (b) at least one process in every read quorum; the write latency is the worse of the two, captured by the outer $\max$. Condition (a) costs $\delta_{\ell M \ell}$, the RTT to the closest majority. For condition (b), each read quorum is satisfied by its member closest to $\ell$ (the $\min$), and the slowest such quorum is latency-determining (the inner $\max$).

\subsubsection{Best Case Read Latency} 
\begin{equation}
    \label{eq:eag-read-blat}
    BC(\Delta_r)(N, p) = \min_{Q \in Q^R}\{ \delta_{pQp} \}
\end{equation}

With no concurrent write, $p$ serves the read as soon as one read quorum replies, so the latency is the RTT to $p$'s closest read quorum, the $\min$ in eq (\ref{eq:eag-read-blat}).

\subsubsection{Worst Case Read Latency}
\begin{equation}
    \label{eq:eag-read-wlat}
    \begin{aligned}
    W&C(\Delta_r(N, p)) = \\
    \min_{Q\in Q^R} \big( \max(&\max_{q\in Q}\{\Delta_w + \delta_{\ell p} - \delta_{\ell q} + \delta_{pq}\}, \\
    &\delta_{pQp})\big)
    \end{aligned}
\end{equation}

\begin{figure}[htbp]
    \centering
    \includegraphics[width=0.44\textwidth]{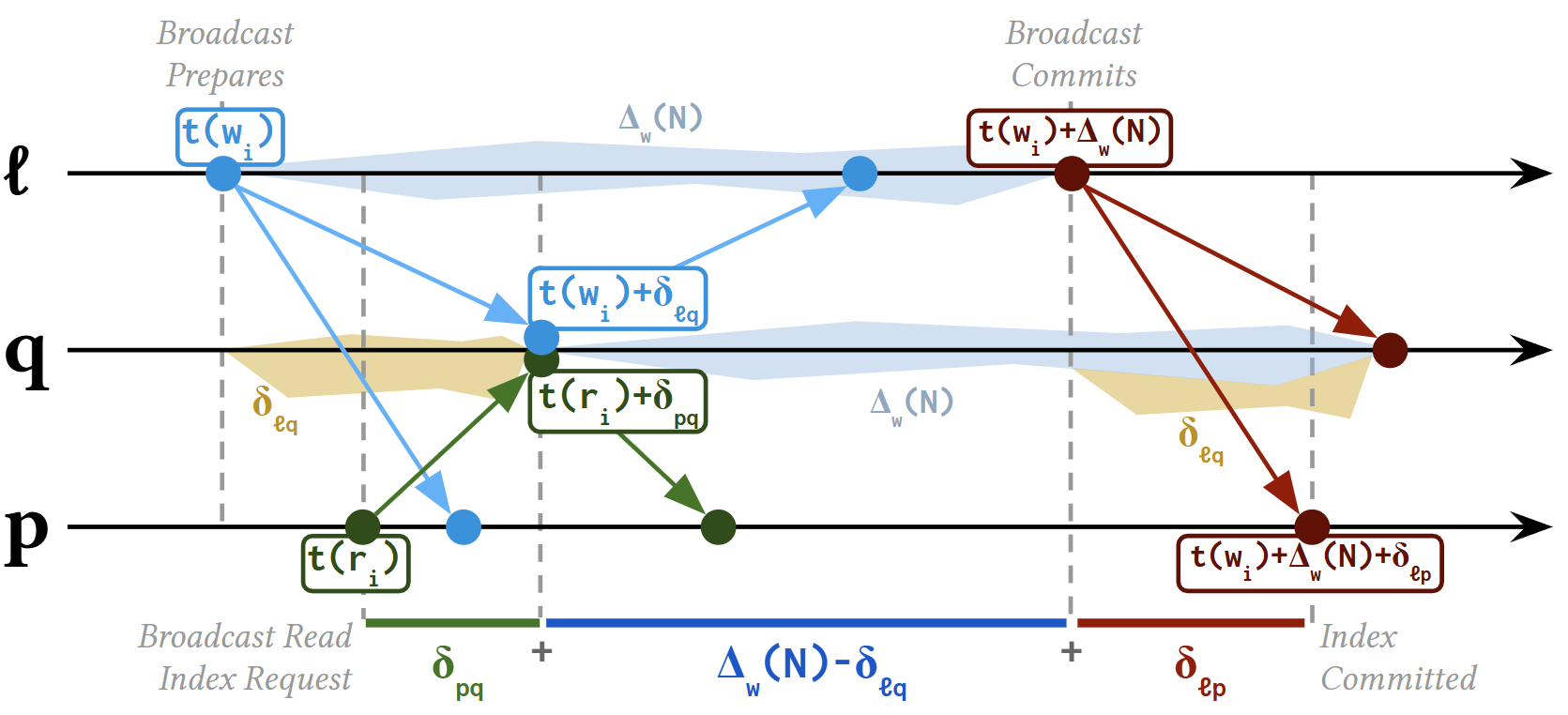}
    \caption{Visualizes the derivation of worst case read latency in eager stamping in the case where read blockage time is the dominating factor}
    \label{fig:n_eager_rlat}
\end{figure}

In eager stamping, $p$ broadcasts for indexes, stamps with the maximum $i_{max}$ its read quorum returns, and serves once $i_{max}$ commits. The worst case is the larger of two costs (outer $\max$): the quorum contact time $\delta_{pQp}$, when the quorum replies so late that $i_{max}$ is already committed at $p$ (case a), or the read blockage time (case b). For case (b), the read blocks longest when the concurrent write $i$ starts as late as possible while still being stamped. To stamp with $i$, it suffices that one $q \in Q$ replies with $i$ to $p$'s request. Taking the worst $q$ (inner-most max), the latest time the write can start is such that $p$'s request reaches $q$ exactly as $q$ receives that prepare, i.e. $t(r_i)+\delta_{pq}=t(w_i)+\delta_{\ell q}$, where $t(r_i)$ and $t(w_i)$ are the read and write invocation times. The read returns when the commit for $i$ reaches $p$, at $t(w_i)+\Delta_w+\delta_{\ell p}$. Subtracting the read invocation time and simplifying gives a latency of $\Delta_w + \delta_{\ell p} - \delta_{\ell q} + \delta_{pq}$. The outer $\min$ then captures $p$'s \textit{closest} quorum, giving the final expression. Figure \ref{fig:n_eager_rlat} illustrates case (b).

\subsection{Invalidation}

Invalidation shares eager stamping's write procedure and best case read latency, so eq (\ref{eq:eag-write-lat}) and eq (\ref{eq:eag-read-blat}) carry over with $L$ substituted for $Q^R$, as invalidation supports only local reads. Its worst case, however, is unbounded.

\subsubsection{Worst Case Read Latency}
\begin{equation}
    \label{eq:inval-read-wlat}
    WC(\Delta_r(N, p)) = \infty
\end{equation}

Invalidation's go event for $i$ fires only when $p$'s latest committed index catches up to its latest received prepare index. Under constant write contention the prepare index is always greater, so the go event may never elapse. Invalidation is therefore  worse than eager stamping, on par without contention, and unbounded with it. We therefore focus on eager versus delayed from here.

\subsection{Delayed Stamping}

Delayed stamping is best of the three mechanisms, but as Figure \ref{fig:n_classification_table_pre} shows it supports only local reads, so we state its differing results in terms of $L$ rather than $Q^R$. Its write procedure matches eager stamping, i.e., the leader attaches a promise time $t+\alpha$ to prepares, and setting $\alpha$ to the RTT to the furthest process the write must wait for makes $t+\alpha$ elapse exactly when the necessary prepares are acknowledged, leaving write latency unchanged (eq \ref{eq:eag-write-lat}). Best case read latency is likewise unchanged (eq \ref{eq:eag-read-blat}). With no concurrent write, all stop and go events have already elapsed and the delayed semantics do not apply.

\subsubsection{Worst Case Read Latency}
\begin{equation}
    \label{eq:del-read-wlat}
    \begin{aligned}
    WC(\Delta_r(N, p)) = \min_{q\in L}\{\max (&\delta_{pq}+\delta_{\ell q}+\delta_{\ell p}, 
    \\&\delta_{pq} + \delta_{qp})\}
    \end{aligned}
\end{equation}

\begin{figure}[htbp]
    \centering
    \includegraphics[width=0.44\textwidth]{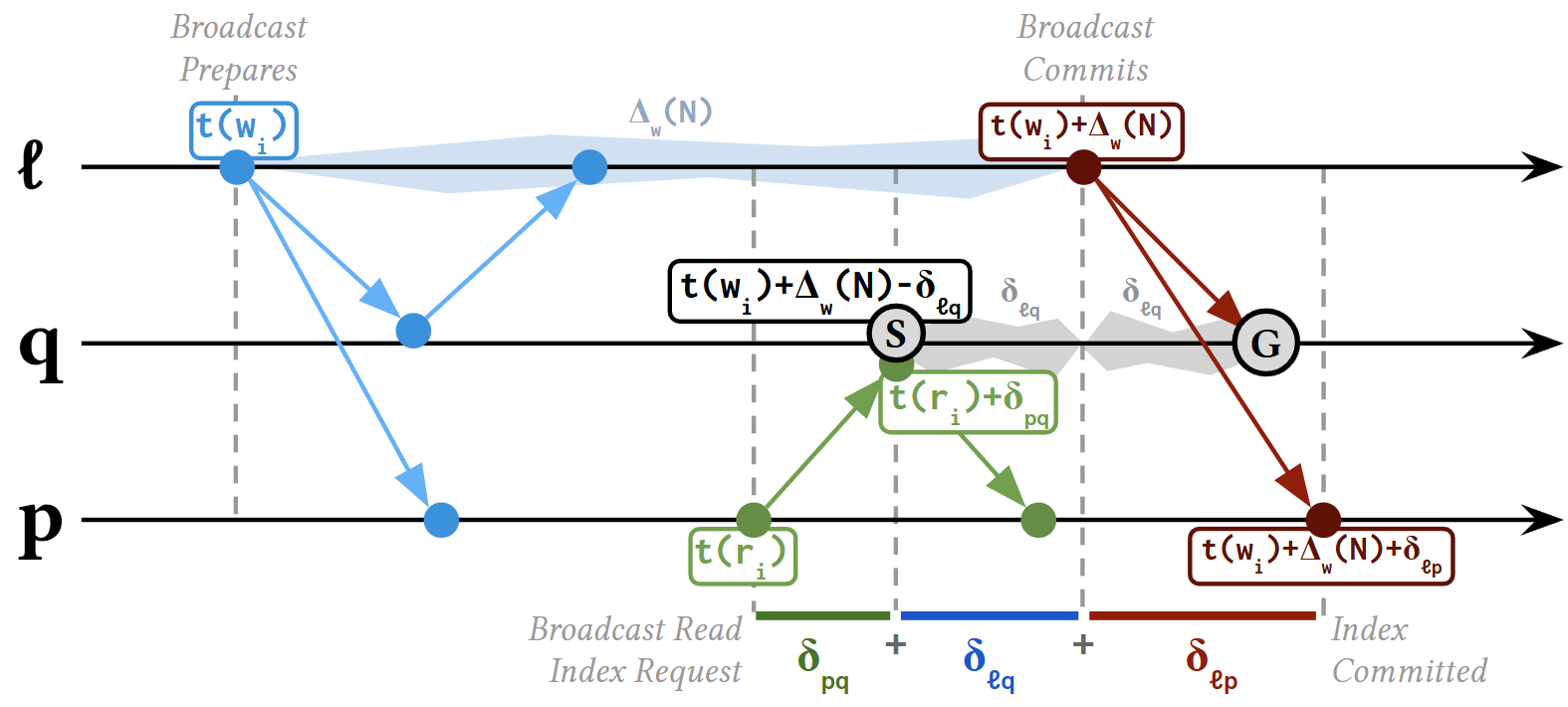}
    \caption{Visualizes the derivation of worst case read latency in delayed stamping in the case where read blockage time is the dominating factor}
    \label{fig:n_delayed_local_rlat}
\end{figure}

Delayed stamping changes only the read blockage term. As before, the worst case is the larger of the contact time to the closest local reader, $\delta_{pq}+\delta_{qp}$ (case a), and the blockage time (case b). The difference is when $q$'s stop event for $i$ elapses. In eager, it is when $q$ receives the prepare, but in delayed it is often later (or at worst equivalent), specifically, half an RTT before the commit, at $t(w_i)+\Delta_w(N)-\delta_{\ell q}$ (Section \ref{subsec:pairwise}, Figure \ref{fig:n_pairwise}). The read blocks longest when $p$'s request reaches $q$ exactly when $q$'s stop event for $i$ elapses, so $t(r_i)+\delta_{pq}=t(w_i)+\Delta_w(N)-\delta_{\ell q}$. The read returns when the commit for $i$ reaches $p$, at $t(w_i)+\Delta_w(N)+\delta_{\ell p}$ and subtracting $t(r_i)$ gives a latency of $\delta_{\ell p}+\delta_{\ell q}+\delta_{pq}$. Unlike eager's term, this contains no $\Delta_w$. Delayed stamping drops the full-write-latency penalty, so blockage shrinks for readers near the leader and is zero at $\ell$ itself. Figure \ref{fig:n_delayed_local_rlat} illustrates case (b).

\subsection{Network Variance}

We model variance as a change in $\delta$, giving a \textit{before network} $N$ and an \textit{after network} $N'$ that differ only in $\delta$, and hence in $M$. The effect of variance on a metric is its difference across the two, for instance $\Delta_w(N') - \Delta_w(N)$ for write latency. Variance occurring during an in-progress operation leaves that operation's metric somewhere between its values for $N$ and $N'$.

Delayed stamping adds one wrinkle. Pairwise's $\alpha$ should always be the RTT to the \textit{furthest} process the leader needs to await, and this RTT changes with variance. Misconfiguring $\alpha$ costs latency but not safety, and the Status Mechanism of \cite{bi2022parameterized} adjusts it automatically. For simplicity, we assume $\alpha$ is always perfectly tuned. In practice, there is a short adjustment period during which a minor performance cost is paid. Additionally, the expressions above heavily rely on the notion of \textit{closest} which is also a function of variance. However, because reads use full broadcast, $p$ always finds its current closest quorum implicitly, so the expressions above stay valid as $\delta$ changes.

\subsection{Other Considerations}
\label{subsec:considerations}

Two factors fall outside our latency analysis but can still guide algorithm choice, and both point to the same escape hatch. First, algorithms that use read leases to support smaller-than-majority read quorums generate passive traffic. Leases are short-lived and must be renewed continually, even when no operations run. Second, leases require an assumption on bounded clock drift, as mentioned in Section \ref{subsec:model}. PQR uses neither leases nor clock assumptions, so deployments that cannot tolerate the background traffic or cannot make clock assumptions are confined to PQR.

\subsection{Trend Analysis}
\label{subsec:categorization-analysis}

We now use these expressions to draw out the trends that motivate our two new algorithms. First, we demonstrate the quorum system's two tradeoff axes from the introduction, which make the optimal configuration depend on the network and workload conditions that shift over time. This is what Ditto (Section~\ref{sec:reconfig}) reconfigures across at runtime to remain adaptive and optimal. Second, we expose the worst case read latency gap in delayed stamping, where wanting a quorum-based configuration currently forces a needless sacrifice in read blockage time, one our Pairwise Quorums generalization (Section~\ref{sec:extending}) removes.

\begin{figure}[htbp]
    \centering
    \includegraphics[width=0.4\textwidth]{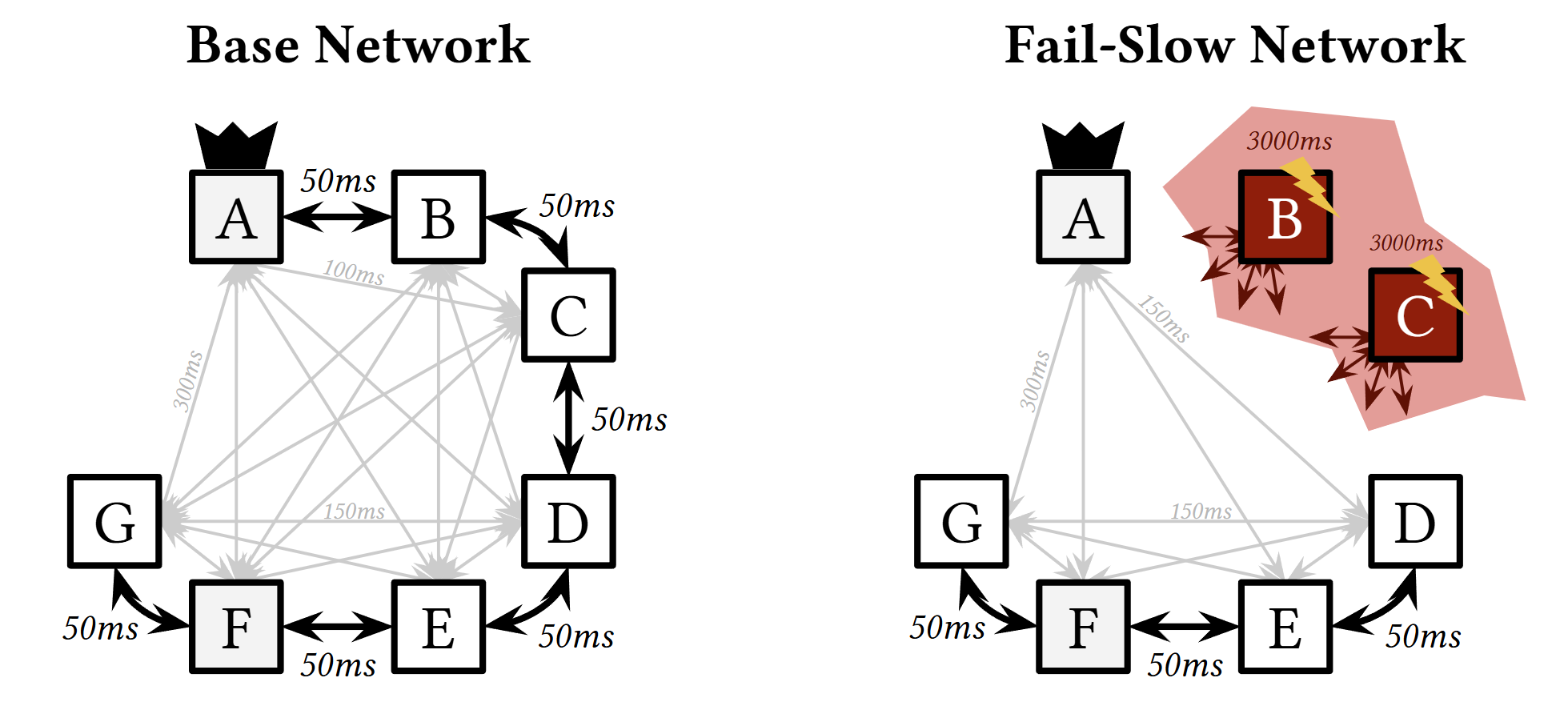}
    \caption{Fully connected ‘line’ network of 7 processes with $A$ as leader, and a version of the same network with fail-slow processes $B$ and $C$}
    \label{fig:n_trends_nets}
\end{figure}

We use a fully connected line network of seven processes $\{A,B,C,D,E,F,G\}$ with leader $A$ and symmetric $50ms$ links, the \textit{Base Network} of Figure \ref{fig:n_trends_nets}. For the heatmaps we will present, we restrict $Q^R$ to two configuration axes reflecting the key tradeoffs, $r = |R|$ and $q = |Q|$, where $Q^R$ is the set of all size-$q$ subsets of the $r$ processes closest to the leader. For instance $r=3, q=2$ gives $Q^R = \{\{A,B\},\{B,C\},\{A,C\}\}$. This omits some configurations such as grid quorums \cite{cheung1992grid} for simplicity of visualization, but the trends we show generalize to arbitrary $Q^R$.

\subsubsection{Static Read-Write Tradeoff} 

\begin{figure}[htbp]
    \centering
    \includegraphics[width=0.29\textwidth]{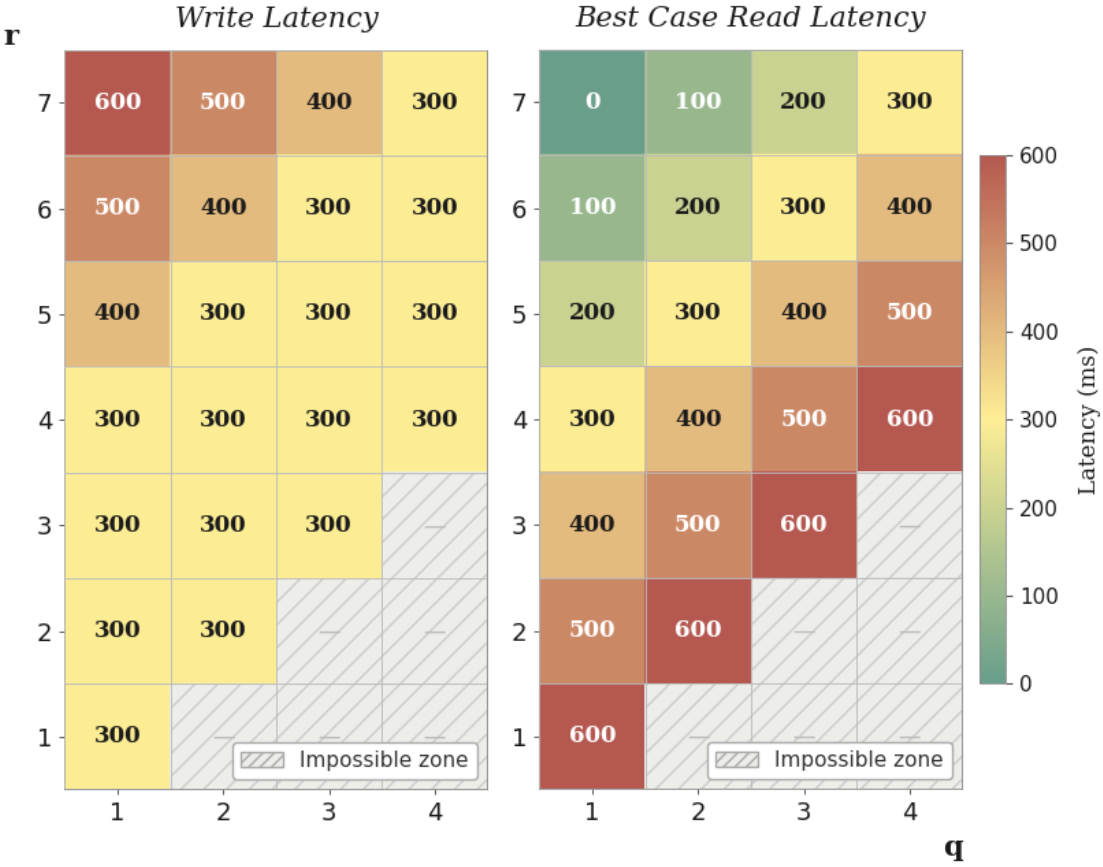}
    \caption{Write latency and best case read latency (read at $G$) from equations (\ref{eq:eag-write-lat}) and (\ref{eq:eag-read-blat}) as a function of $Q^R$ defined in terms of $r$ and $q$ (same for delayed and eager).}
    \label{fig:n_trends_wlat_bcrlat}
\end{figure}

Figure \ref{fig:n_trends_wlat_bcrlat} plots write latency and best case read latency over all $r,q$ following from the mathematical expressions, with reads at $G$ (the two are equal for eager and delayed in the best case). The two metrics trend oppositely. Read latency falls as $r$ grows or $q$ shrinks. At $r=7, q=1$, $G$'s closest quorum is itself with zero latency, whereas at $q=2$ it becomes $\{F,G\}$, adding a $100ms$ RTT; lowering $r$ likewise pushes $G$'s closest quorum away. Write latency moves the other way, floored at $300ms$ by the SMR majority requirement. The key point is that the $q=1$ column, local reads, already spans the entire read-write latency tradeoff and is the only configuration reaching zero read latency; larger $q$ only adds redundancy. So where network variance is negligible and the considerations of Section \ref{subsec:considerations} do not apply, local reads suffices.

This is the first fundamental tradeoff of quorum systems, generalizing to arbitrary $Q^R$: shrinking $R$ towards the leader while keeping fixed quorum sizes trades write latency for read latency and vice versa. A read-heavy workload wants one end and a write-heavy one the other.

\subsubsection{Network Variance Tradeoff} 

\begin{figure}[htbp]
    \centering
    \includegraphics[width=0.29\textwidth]{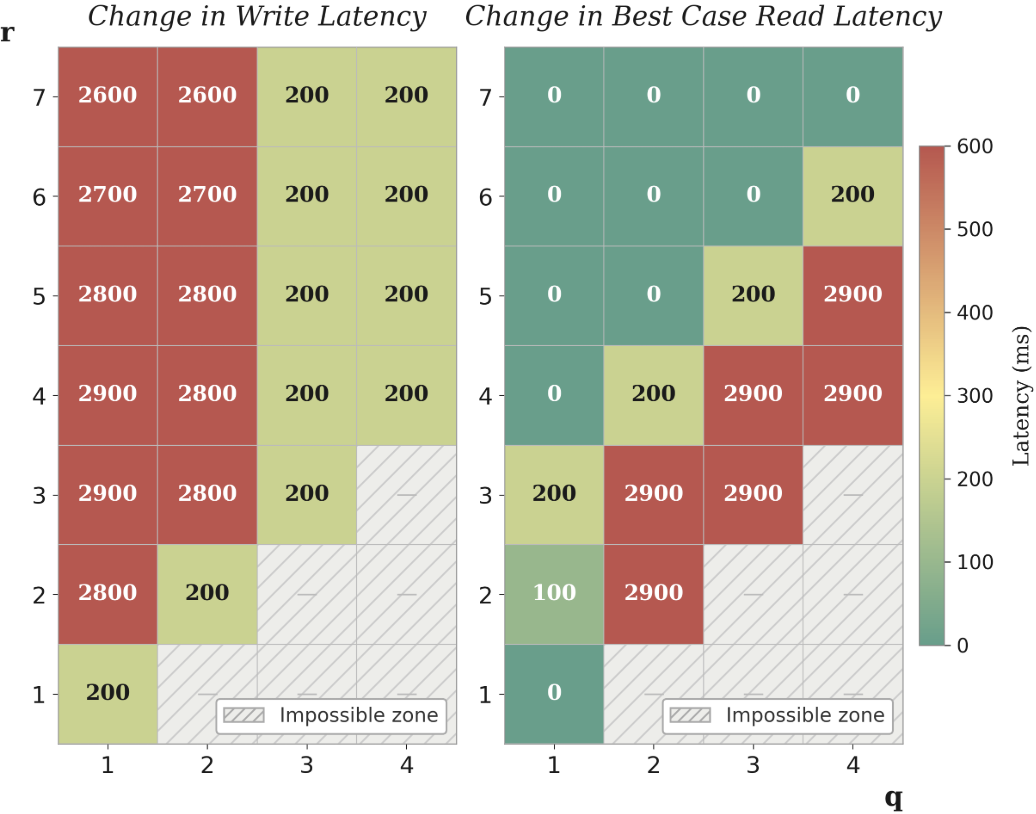}
    \caption{The change in write latency and best case read latency (read at $G$) after spiking processes $B$ and $C$.}
    \label{fig:n_trends_var_wlat_bcrlat}
\end{figure}

To see where local reads breaks down, we spike $B$ and $C$ to be fail slow \cite{gunawi2018fail} by adding $3000ms$ to all of their incoming links, as reflected in the \textit{Fail-Slow Network} of Figure \ref{fig:n_trends_nets}, leaving the leader fast (a slow leader would force the spike on every write regardless of configuration, calling for a leader change instead). Figure \ref{fig:n_trends_var_wlat_bcrlat} plots the resulting change in write and best case read latency from before the spike to after. For writes, low-$q$ configurations, especially local reads, must contact a slow process to satisfy quorum intersection and absorb the spike as a result, while larger $q$ gives the leader enough redundancy to route around the slow processes. All configurations still take an unavoidable $200ms$ increase, since the leader must now use the farther $E$ and $F$ for its majority. For reads, the full spike appears only when every read quorum contains a slow process, so reads are most tolerant at $q=1, r=7$ (Pairwise), where a fast local reader always exists. 

This is the second tradeoff of quorum systems, again generalizing to arbitrary $Q^R$: at fixed $R$, growing quorum sizes trades read latency for write tolerance to network variance. High variance wants one end, low variance the other.

\smallskip

Together these two axes show that the full space of quorum configurations matters, with different conditions calling for different points. This is what Ditto reconfigures across at runtime, keeping a system optimal as conditions change.

\subsubsection{Worst Case Read Latency Gap} 

\begin{figure}[htbp]
    \centering
    \includegraphics[width=0.43\textwidth]{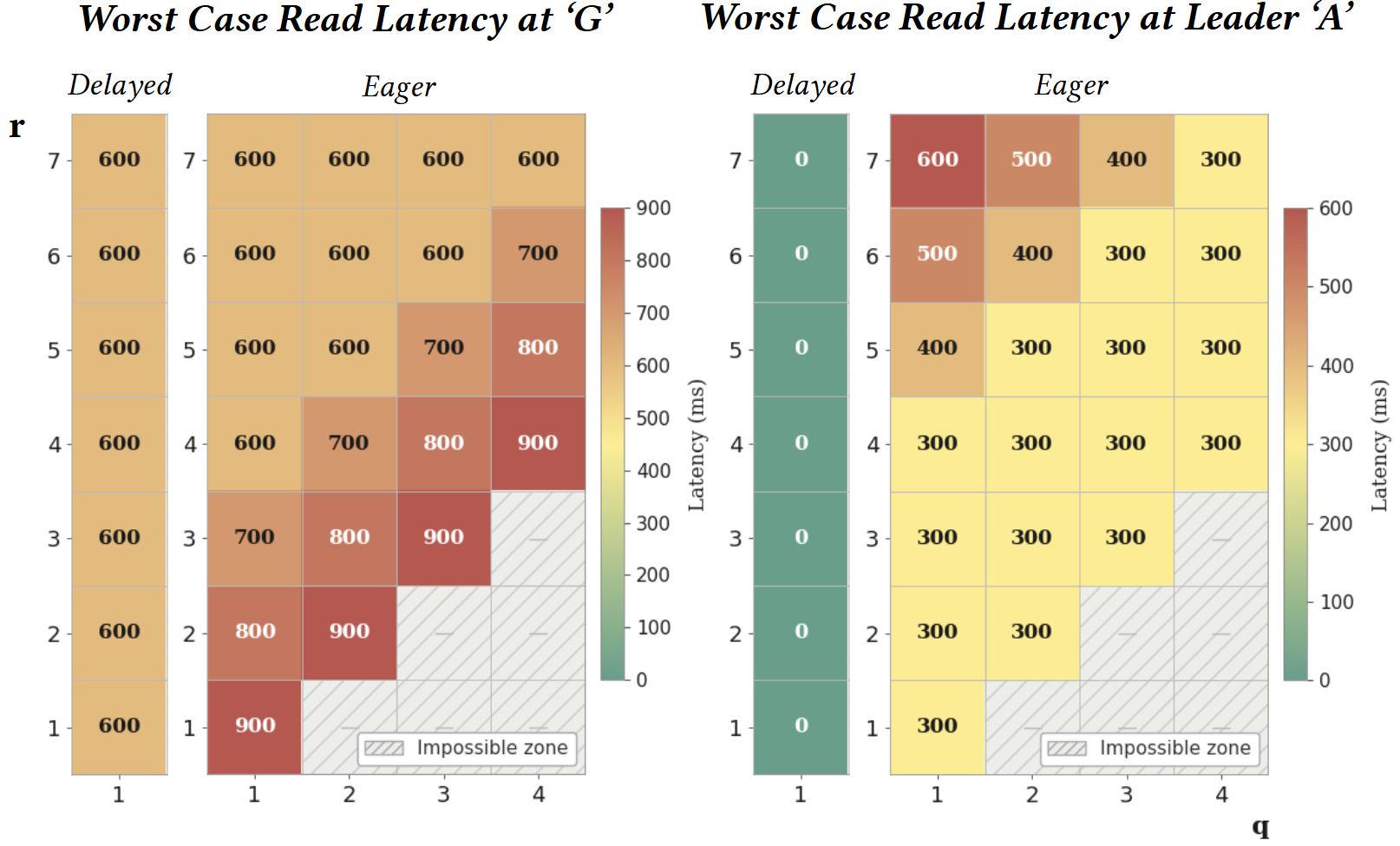}
    \caption{Worst case read latency heatmaps at leader $A$ and at $G$, comparing eager and delayed stamping (delayed stamping only supported for local reads).}
    \label{fig:n_trends_wcrlat}
\end{figure}

Finally, Figure \ref{fig:n_trends_wcrlat} shows worst case read latency, under maximal write contention, for reads at the leader $A$ and at $G$, placing delayed stamping's $q=1$ column beside the full eager heatmaps. Delayed stamping improves the worst case almost everywhere, and dramatically near the leader: $A$ sees zero blockage under delayed against the full write latency under eager, while the distant $G$ sees a smaller gain. The gain is confined to local reads due to the gap in delayed stamping. A deployment that wants write tolerance to network variance needs a quorum-based ($q>1$) configuration, which forces it onto eager stamping and its full read blockage. Closing this gap, extending delayed stamping to arbitrary quorums $Q^R$, is what our algorithm Pairwise Quorums (Section \ref{sec:extending}) solves.

\section{Pairwise-Quorums Algorithm}
\label{sec:extending}

We now present Pairwise Quorums, which closes the delayed stamping gap shown in Figure \ref{fig:n_trends_wcrlat} by extending delayed stamping to arbitrary quorum configurations. Its write procedure is identical to Pairwise's \cite{thiessen2025asymmetric} (Section \ref{subsec:pairwise}) except generalized to quorums (i.e \ref{subsubsec:general_write_alg}), so we focus on the read procedure, building on the general read procedure of Section \ref{subsubsec:general_read_alg}. To focus on the stamping mechanism, we assume every process knows a static $Q^R$.

\subsection{Read Procedure}
In standard Pairwise every read quorum is a single local reader. Following the general read procedure, $p$ contacts its closest local reader, which returns the highest index $i$ for which it has received a prepare and its stop event has elapsed. With a single-process quorum this index is already the stamp, so no aggregation is needed; $p$ blocks until $i$ commits and its own go event elapses, then returns.

\begin{algorithm}[t]
\small
\DontPrintSemicolon

\Init(\textbf{procedure} \text{read():}){
    $cntr \coloneqq cntr + 1$   \tcc*[f]{Local operation identifier}

    \textbf{send} $\langle READ, cntr \rangle$ to all processes

    $replyMap \coloneqq \emptyset$  \tcc*[f]{map index to set of SNP/SP/NA}

    $repliedProcs \coloneqq \emptyset$
    
    \Init(\textbf{upon receiving} $\langle R\_ACK, cntr, M \rangle$ from $q$:){
        $repliedProcs \coloneqq repliedProcs \cup \{q\}$

        \ForEach{$index$ in $M.\text{keys}$}{
            $replyMap[index] = replyMap[index] \cup M[index]$
        }
    }
    
    \textbf{wait for} $\exists Q \in Q^R, \forall w \in Q, w \in repliedProcs$

    \tcc{Stamp with max index with an SP reply and no SNPs}
    $stampIdx \coloneqq \max\{i \mid i \in replyMap.keys() \wedge SP \in replyMap[i] \wedge  SNP \notin replyMap[i]\}$

    \textbf{wait for} $stampIdx \text{ is } \textbf{committed}$
    
    \textbf{wait for} $go\_event[stampIdx] < \text{current\_time()}$
    
    \textbf{return} $stateMachine.read(stampIdx)$
}

\BlankLine

\Init(\textbf{upon receiving} $\langle READ, c \rangle$ from $p$:){
    \tcc{Compute per-index reply map with SP/SNP/NA}
    $IndexMap \coloneqq \emptyset$
    
    \ForEach{i \textbf{in} 0..vidx}{
        \tcc{Prepare not received (no stop event)}
        \lIf{prepare for $i$ not received}{ 
        $IndexMap[i] \coloneqq \text{NA}$ }

        \tcc{stop event for i not passed}
        \lElseIf{$\text{current\_time()} < stop\_event[i]$}{ $IndexMap[i] \coloneqq \text{SNP}$ }

        \tcc{stop event for i passed}
        \lElse{ $IndexMap[i] \coloneqq \text{SP}$ }
    }

    \textbf{send} $\langle R\_ACK, c, IndexMap \rangle$ to $p$
}

\BlankLine

\caption{Read pseudocode for Pairwise Quorums assuming no message loss, all processes are non-faulty, a fixed leader $\ell$, and a known $Q^R$}
\label{alg:read_procedure}
\end{algorithm}

\begin{figure}[htbp]
    \centering
    \includegraphics[width=0.45\textwidth]{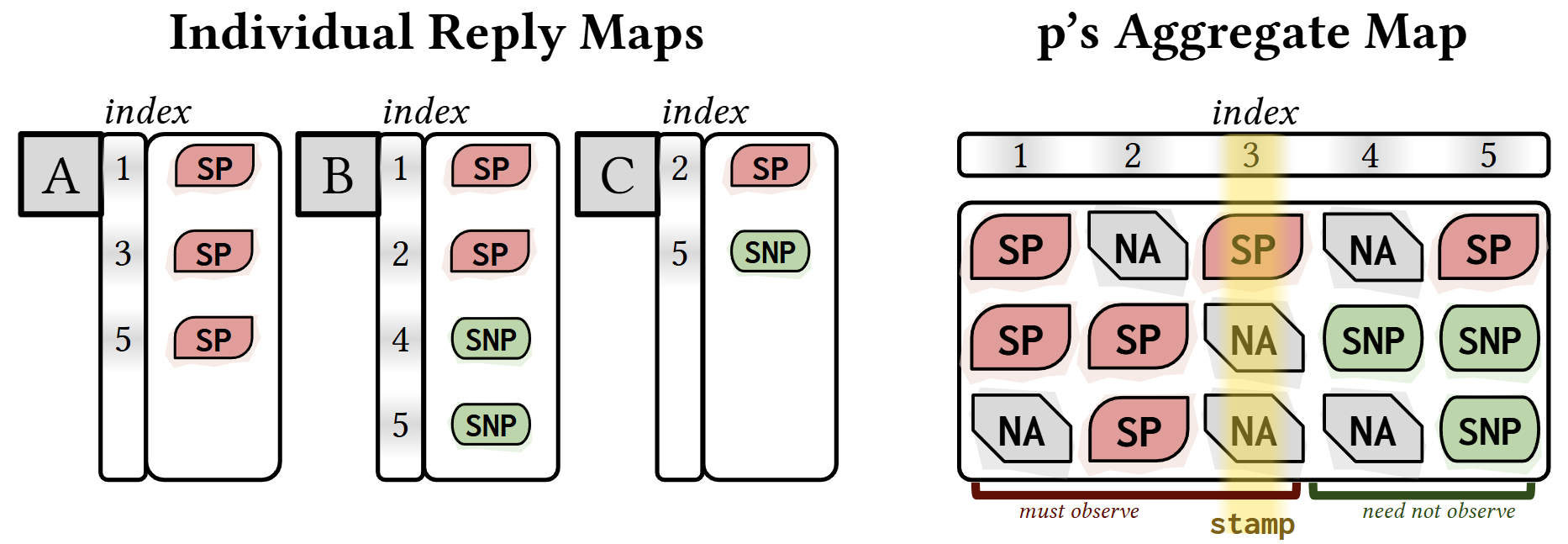}
    \caption{Example of Pairwise Quorums stamping mechanism with process $p$ aggregating stop event information from quorum $\{A, B, C\}$ and stamping with index 3.}
    \label{fig:n_pairwise_quorums_demo}
\end{figure}

With larger read quorums, no single quorum member can decide the stamp alone. Each member $w$ instead replies with a per-index status map: \textbf{SP} (the stop event for that index has elapsed), \textbf{SNP} (a prepare was received but the stop event has not elapsed), or \textbf{NA} (no prepare was received). $p$ aggregates these maps and stamps with the highest index that has at least one SP and no SNP (Algorithm \ref{alg:read_procedure}). The reasoning is that an SP means that the member has entered its read blockage interval for that index and cannot be certain whether the write finished, so the read must conservatively observe it. A single SNP, on the other hand, proves the write has not finished, since every stop event elapses before the promise time, so $p$ may safely skip the index. Figure \ref{fig:n_pairwise_quorums_demo} shows an example execution in which $p$ contacts $\{A, B, C\}$ and stamps with index 3.

A natural concern is that reply size grows with the number of writes, since each member returns a per-index map. However, in practice only indexes between the latest committed and the latest received prepare are uncertain; anything earlier is already committed and must be observed. Therefore, members include only this range, bounding the reply map size by the number of in-progress writes.

\subsection{Correctness}

We sketch correctness through the four real-time ordering cases of linearizability.

\textit{Write before read.} If write $o_w$ completes before read $o_r$ starts, $o_r$ must observe $o_w$. Say $o_w$ committed at index $i$; then $o_r$ must stamp with at least $i$. Suppose $o_r$ stamped using quorum $Q$. Since $o_w$ completed, every process's stop event for $i$ has passed, and by the intersection requirement at least one $w \in Q$ received a prepare for $i$. So $p$'s aggregate map holds at least one SP for $i$ and no SNP, thus $p$ stamps with at least $i$.

\textit{Write before write.} Unchanged from Pairwise \cite{thiessen2025asymmetric}, since we keep its write procedure.

\textit{Read before write.} If $o_r$ completes before $o_w$ starts, $o_w$'s read-modify-write must observe at least the state read by $o_r$. If $o_r$ stamped with $i$, then $i$ committed at $p$ and so completed, and by the write-before-write case $o_w$ observes $i$.

\textit{Read before read.} If $o_{r1}$ completes before $o_{r2}$ starts, $o_{r2}$ must observe at least the state read by $o_{r1}$. If $o_{r1}$ stamped with $i$, then $i$ committed and so completed, and by the write-before-read case $o_{r2}$ stamps with at least $i$.

\subsection{Analysis}
\label{subsec:extending-analysis}

We now fill the gap from Section \ref{sec:categorization} by extending the delayed stamping worst case read latency, eq (\ref{eq:del-read-wlat}), from $L$ to arbitrary $Q^R$.

\subsubsection{Worst Case Read Latency}
\begin{equation}
    \label{eq:del-read-wlat-gen}
    \begin{alignedat}{2}
    WC(\Delta_r(N, p)) &= \min_{Q \in Q^R}\Big( \max \big\{\\
    \max_{q \in A(Q,p)} &\{ \delta_{pq} + \delta_{\ell p} + \min(\delta_{\ell q}, \Delta_w(N)-\delta_{\ell q})\}, \\
    &\qquad \delta_{pQp} \big\} \Big) \\
        \text{where} \qquad \qquad \quad  \\
        A(Q, p) = \{q_1 \in &Q\mid \forall q_2 \in Q, \\
    \neg (\delta_{\ell q_2} &< \max(\Delta_w(N) - \delta_{\ell q_1}, \delta_{\ell q_1}) - \delta_{pq_1} + \delta_{pq_2} \\
      & < \Delta_w(N) - \delta_{\ell q_2})\}
    \end{alignedat}
\end{equation}

The outer $\min$ picks the closest quorum $Q$, and within it the latency is the worse of (a) the quorum contact time $\delta_{pQp}$, identical to case (a) of eq (\ref{eq:eag-read-wlat}), and (b) the read blockage time, the max over $q \in A(Q,p)$.

For case (b), we find the latest time a concurrent write $i$ can start while the read still stamps with $i$. The protocol needs at least one $q \in Q$ to reply SP, so $q$ must have received the prepare for $i$ and its stop event elapsed. Thus the latest time the write can start is when $p$'s request reaches $q$ at the exact first moment $q$ can reply SP, namely $t(r_i)+\delta_{pq} = \max(t(w_i) + \delta_{\ell q},\; t(w_i) + \Delta_w(N) - \delta_{\ell q})$, the later of the prepare arriving and the stop event elapsing. We call this \textit{aligning to} $q$; the prepare term dominates only for processes farther than the promise time accounts for, those in large quorums the leader need not intersect. The read returns when the commit reaches $p$, at $t(w_i) + \Delta_w(N) + \delta_{\ell p}$. Subtracting $t(r_i)$ via the alignment equality and applying the identity $\Delta_w -\max(\Delta_w-\delta_{\ell q}, \delta_{\ell q})=\min(\Delta_w - \delta_{\ell q}, \delta_{\ell q})$ gives $\delta_{\ell p} + \delta_{pq} + \min(\delta_{\ell q}, \Delta_w(N) - \delta_{\ell q})$, the inner expression of eq (\ref{eq:del-read-wlat-gen}).

\begin{figure}[htbp]
    \centering
    \includegraphics[width=0.25\textwidth]{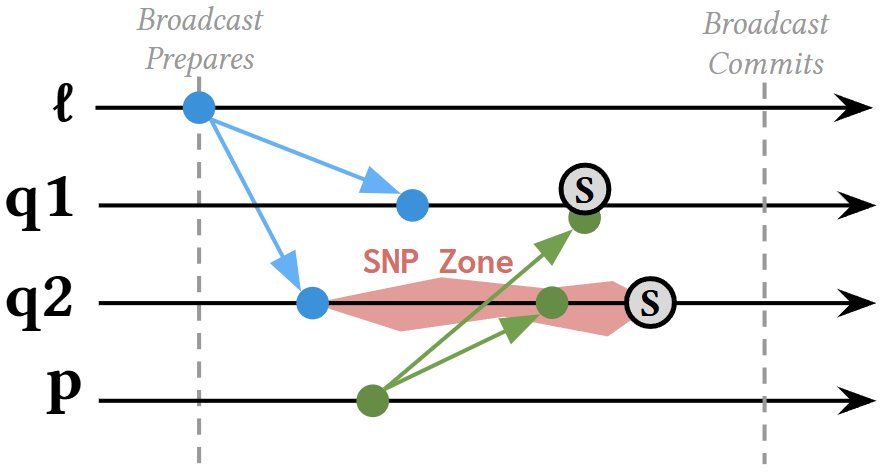}
    \caption{\textit{Aligning} to $q_1$ allows the write to start the \textit{latest}, however, $q_2$ would end up returning SNP making this the incorrect worst case.}
    \label{fig:n_snp_zone}
\end{figure}

We now address $A(Q, p)$. With quorums larger than one, aligning to some $q_1$ may push another $q_2 \in Q$ into its \textit{SNP zone}, the interval between when $q_2$ receives the prepare for $i$ and when its stop event elapses, during which $q_2$ replies SNP and prevents stamping with $i$ (Figure \ref{fig:n_snp_zone}). $A(Q, p)$ is therefore the set of \textit{alignable} $q_1 \in Q$: those that when \textit{aligned to} leave every other $q_2$ outside its SNP zone. Writing that condition for $q_2$, $\neg(t(w_i) + \delta_{\ell q_2} < t(r_i) + \delta_{pq_2} < t(w_i) + \Delta_w(N) - \delta_{\ell q_2})$, and substituting the alignment equality for $t(r_i)$ gives the set builder of eq (\ref{eq:del-read-wlat-gen}). This makes the worst case expression tight by excluding seemingly valid but actually impossible constructions.

\begin{figure}[htbp]
    \centering
    \includegraphics[width=0.3\textwidth]{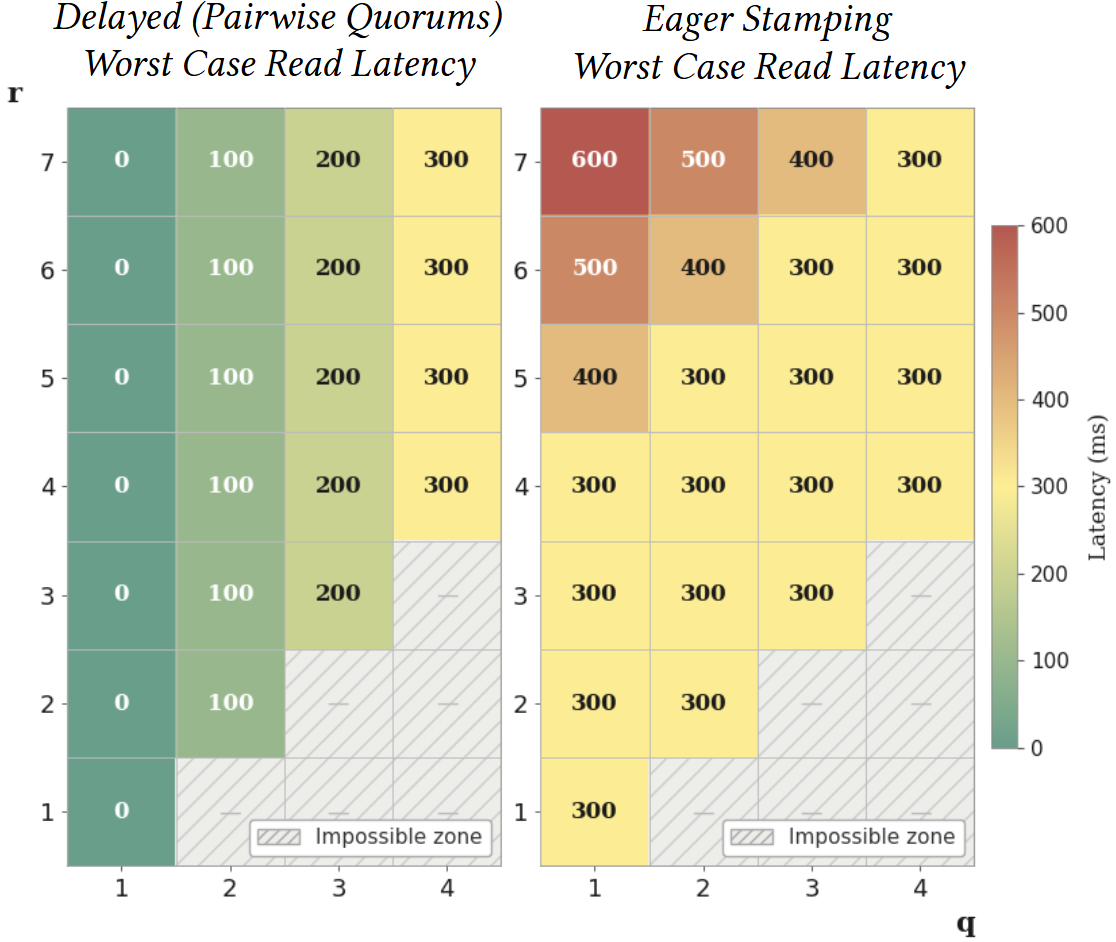}
    \caption{Worst case read latency compared analytically between Pairwise Quorums and eager stamping for reads at leader $A$ in the same \textit{Base Network}, showing generalization of read blockage time improvement in delayed stamping.}
    \label{fig:n_del_full_heatmap}
\end{figure}

Figure \ref{fig:n_del_full_heatmap} plots worst case read latency for Pairwise Quorums against eager stamping at leader $A$ in the \textit{Base Network}. The delayed stamping improvement we observed initially only for local reads generalizes across every quorum-based configuration, so deployments can now tune both tradeoff axes without giving up reduced read blockage time.

\section{Ditto Algorithm for Reconfiguration}
\label{sec:reconfig}

We now present Ditto, our mechanism for reconfiguring $Q^R$ at runtime. Ditto reconfigures at runtime and builds on Pairwise Quorums thus retaining delayed stamping's reduced read blockage at every configuration. Under our model, Ditto subsumes every existing algorithm: against each, it is either better, using delayed stamping where the other uses eager or invalidation, or reconfigures to match it on the tradeoff axes where no single configuration dominates.

Ditto stores $Q^R$ in a blackbox linearizable object, the \textit{configuration layer}, as an explicit set of sets. Every write and every read first reads this layer to learn the current $Q^R$. A write does so to know which processes it must intersect, and a read, to know which quorums it may contact.

We require two things of the configuration layer. First, it must make progress under up to $f$ faults in the main layer, so that a faulty quorum can always be reconfigured out of $Q^R$ to avoid stalling writes. Second, non-faulty processes must read it with zero latency, so that consulting it adds no overhead to a main-layer operation. We satisfy both by running the configuration layer as a second MultiPaxos instance on the same processes and leader, using Pairwise with every process a local reader. This gives zero-latency configuration reads everywhere, at the price of configuration writes contacting every process, following from the read-write tradeoff our analysis described earlier. Because the configuration layer runs Pairwise, it already handles all fault-tolerance through leases and inherits a fault tolerance of $f$, so the main layer itself uses no leases. If a read quorum becomes faulty on the main layer, a write to the configuration layer is performed to reconfigure out the faulty quorum, and all lease and fault tolerance mechanisms are handled there.

Reconfiguration can violate linearizability if handled naively. Consider $\{A,B,C,D,E\}$ with $Q^R = \{\{B\}\}$ and writes committed through index $i=10$. Suppose the leader satisfied intersection using $\{A,B,C\}$ and its prepares never reached $D$ or $E$. Reconfiguring to $Q^R = \{\{D\}\}$ now lets a read contact $D$, which stamps with $i=0$ and misses all 10 committed writes.

The new quorum was never intersected by the earlier writes, so it lacks their indexes. Correctness requires that every new quorum serve reads only once it is at least as up-to-date as some old quorum, that is, it would stamp with an index at least as large. Intuitively, the old quorums must ``pass the torch'' to the new ones.

We enforce this with a \textit{lower bound index} (or ``wall'' index). On each read, $p$ compares the configuration layer's current index against the one it last saw; if they differ, a reconfiguration has occurred, so $p$ fetches the latest prepare index $i_{lb}$ from the leader and stamps with at least $i_{lb}$ for the current and all future reads regardless of what its read quorum returns. Since $i_{lb}$ is at least what any former quorum would have stamped with, the new quorum's possibly stale indexes no longer threaten correctness.

\begin{figure}[htbp]
    \centering
    \includegraphics[width=0.5\textwidth]{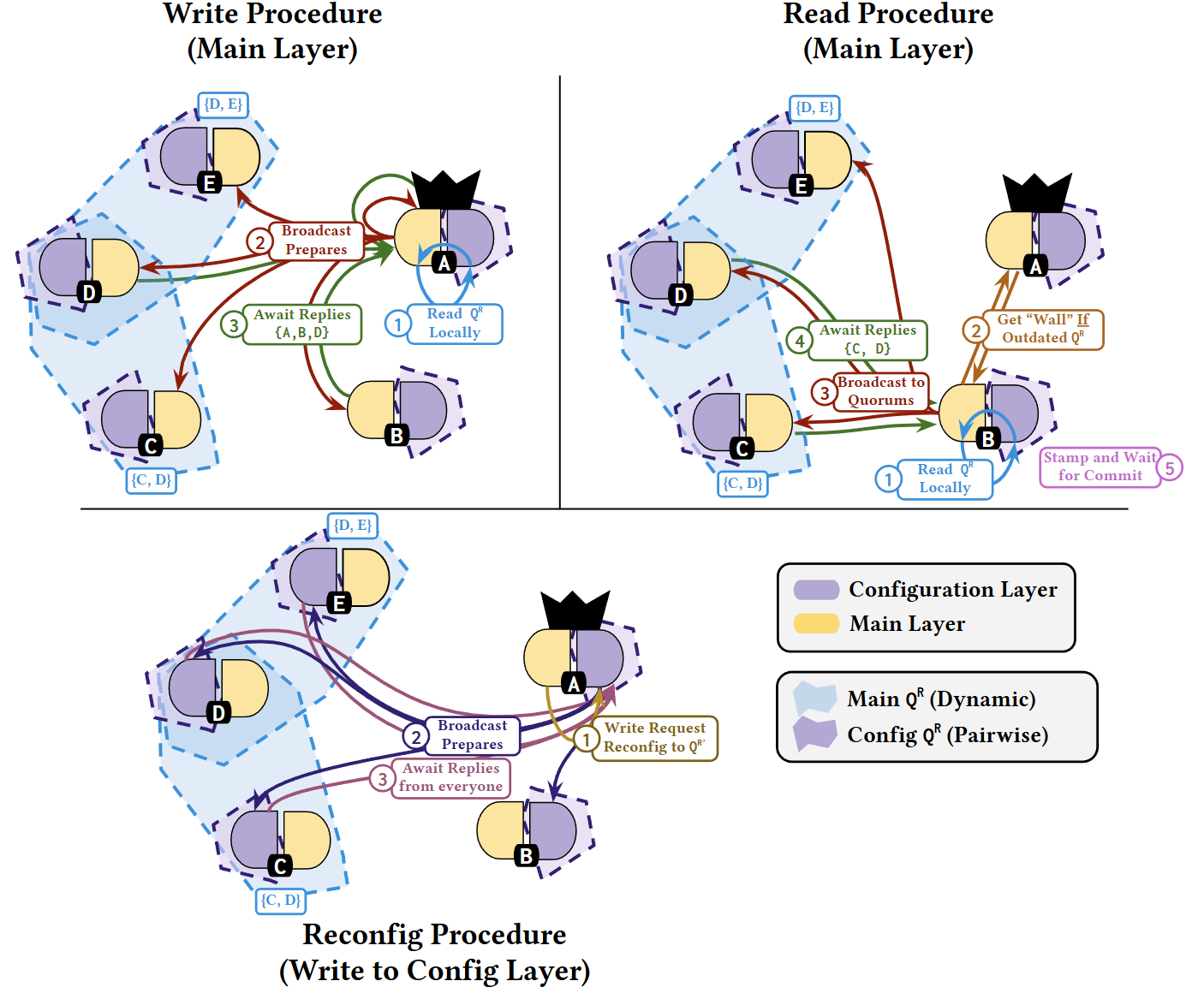}
    \caption{All procedures of the Ditto reconfiguration algorithm visualized.}
    \label{fig:n_ditto_algo}
\end{figure}

The read, write, and reconfiguration procedures are illustrated in Figure \ref{fig:n_ditto_algo}.

Ditto's overhead is modest: a zero-latency configuration read on every operation, and, only while a reconfiguration is in flight, a full config-layer write plus a short window where reads pay an extra round trip to the leader for $i_{lb}$ and see elevated read blockage. These costs buy the ability to move $Q^R$ across the two tradeoff axes at runtime, keeping a system at the optimal configuration as its workload and network change.

\section{Evaluation}
\label{sec:experiments}

Our evaluation has two goals. First, we validate the analytical expressions of Sections \ref{sec:categorization} and \ref{sec:extending} against a real implementation. Second, we demonstrate the two practical benefits the analysis predicts: that Pairwise Quorums removes the forced tradeoff between stamping mechanism and tolerance to network variance, and that Ditto reconfigures to remain optimal as conditions change. We first reproduce the analytical heatmaps experimentally, then demonstrate both benefits on an emulated geo-distributed network.

\subsection{Experimental Setup}

\textbf{Implementation.} Our prototype runs on the JVM and realizes the full framework of Section \ref{sec:categorization}: every existing algorithm, as well as Pairwise Quorums and Ditto, is obtained by selecting a stamping mechanism and a read-write quorum system, so all baselines and our own algorithms execute on identical code paths and differ only in configuration.

\textbf{Testbeds.} We use two networks, both realized on separate virtual machines with per-link latencies enforced by Linux traffic control. The first is the emulated \textit{Base Network} of Figure \ref{fig:n_trends_nets}: seven processes on a fully connected line with symmetric $50$ms links, together with its \textit{Fail-Slow} variant in which processes $B$ and $C$ are spiked to $3000$ms. The second is an emulated geo-distributed network of five processes whose pairwise latencies are set to measured AWS p50 inter-region round-trip times across \texttt{ca-central-1} (the leader), \texttt{us-east-1}, \texttt{us-west-1}, \texttt{eu-north-1}, and \texttt{eu-central-1}.

\textbf{Methodology.} Latency is measured from the moment a process receives a request to when it is about to return it. Each experiment dispatches a set number of operations at a set dispatch interval, run concurrently (exact values stated per experiment). Because the JVM introduces latency through just-in-time compilation and garbage collection that the analytical model does not capture, we warm up the JVM before measuring and discard outliers. We also set the Pairwise $\alpha$ promise time to $10ms$ above the RTT to the furthest process, penalizing writes by $10ms$ but keeping reads optimal -- a suggested buffer for real deployments.

\subsection{Validating the Analytical Model}

\begin{figure}[htbp]
    \centering
    \includegraphics[width=0.33\textwidth]{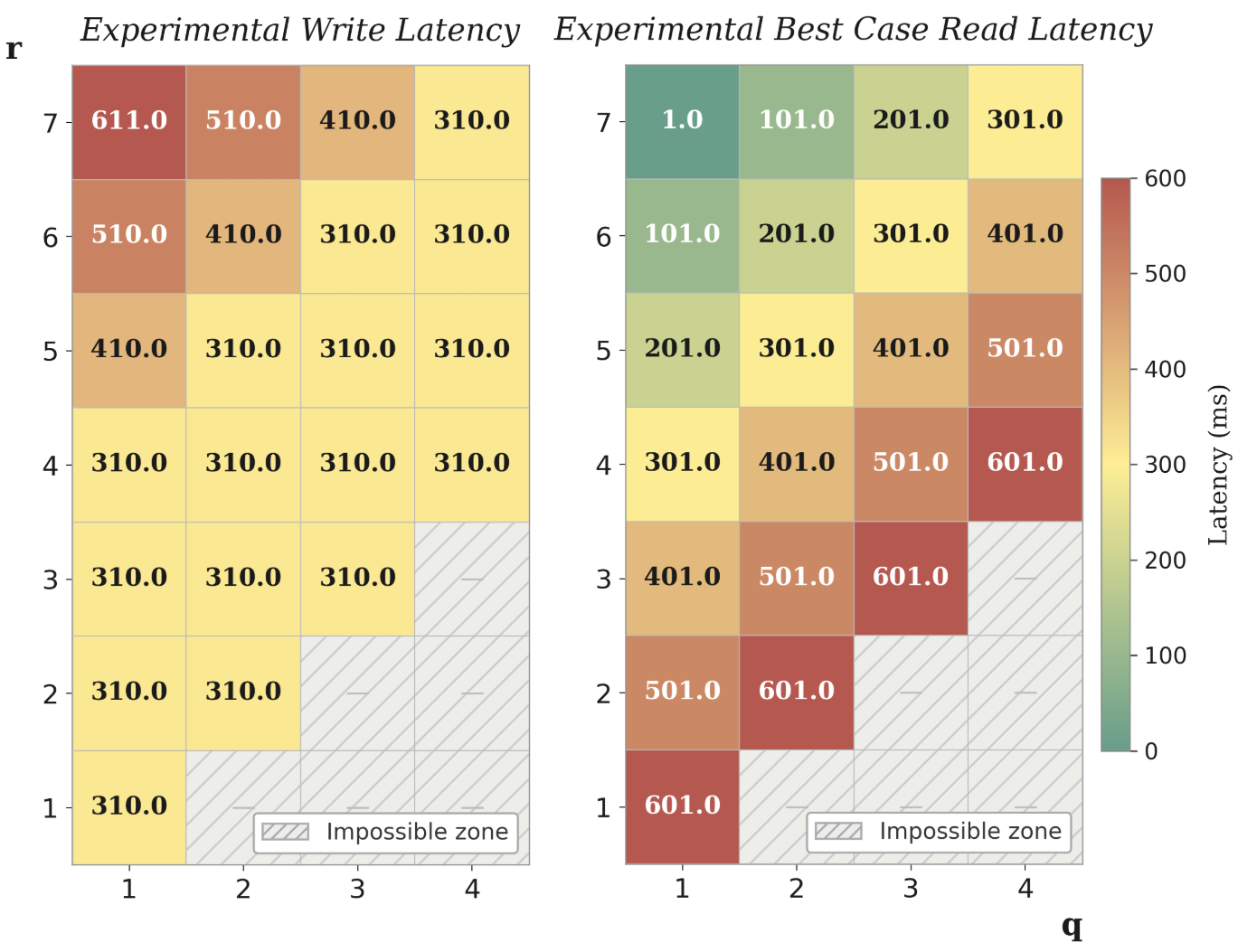}
    \caption{Measured write and best case read latency (reads at $G$) over the $(r,q)$ space on the Base Network, reproducing Figure \ref{fig:n_trends_wlat_bcrlat}.}
    \label{fig:n_eval_bcrlat_wlat_repeat}
\end{figure}

Figure \ref{fig:n_eval_bcrlat_wlat_repeat} shows measured write and best case read latency across the $(r,q)$ space on the \textit{Base Network}, with reads executing at process $G$. We realize each configuration with EdgePQR, which supports the full quorum space under eager stamping; since write latency and best case read latency are identical across stamping mechanisms (Section \ref{sec:categorization}), this validates those expressions for every configuration. For each cell, we execute $5000$ operations at a $2ms$ dispatch interval and report a trimmed mean over the interquartile range, discarding the top and bottom quartiles to remove compilation and collection outliers. The write heatmap runs an all-writes workload, and the read heatmap an all-reads workload with reads at $G$ and no concurrent writes. The measured heatmaps reproduce the analytical ones of Figure \ref{fig:n_trends_wlat_bcrlat}, deviating by at most a few milliseconds per cell, attributable to JVM scheduling, measurement noise, and the $10ms$ $\alpha$ buffer.

\begin{figure}[htbp]
    \centering
    \includegraphics[width=0.33\textwidth]{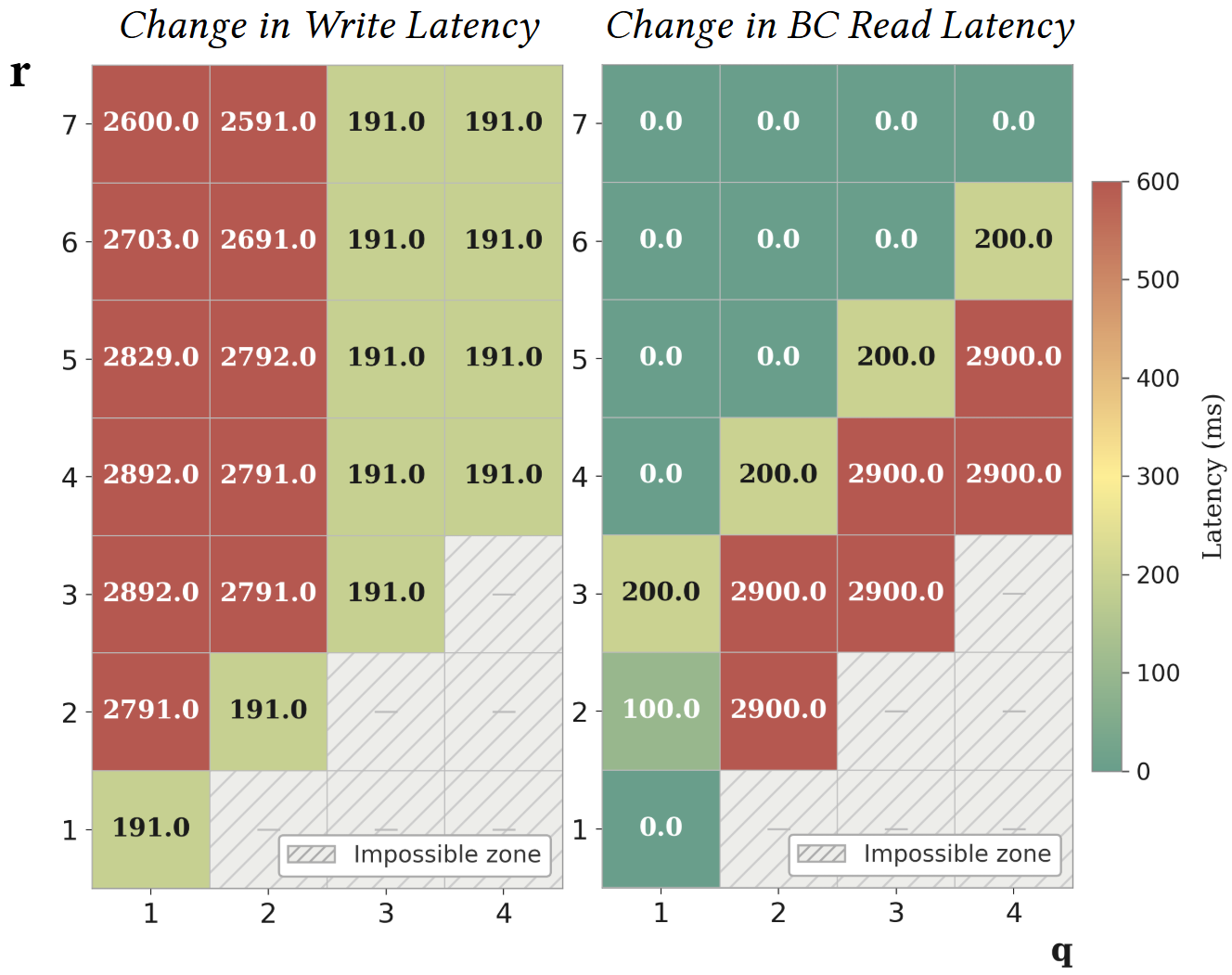}
    \caption{Measured change in write and best case read latency (reads at $G$) over the $(r,q)$ space from the Base to the Fail-Slow Network, reproducing Figure \ref{fig:n_trends_var_wlat_bcrlat}.}
    \label{fig:n_eval_variance_repeat}
\end{figure}

Figure \ref{fig:n_eval_variance_repeat} measures how write and best case read latency change between the Base Network and the Fail-Slow Network, again reading at $G$. We rerun the previous experiment on both networks and report the per-cell difference after interquartile trimming. We raise the dispatch interval to $7.5ms$ here because at $2ms$ the Fail-Slow Network cannot drain operations fast enough, and network congestion rather than the fail-slow spike becomes the dominant slowdown. The measurements again match the analytical heatmaps of Figure \ref{fig:n_trends_var_wlat_bcrlat} to within a few milliseconds per cell.

\begin{figure}[htbp]
    \centering
    \includegraphics[width=0.33\textwidth]{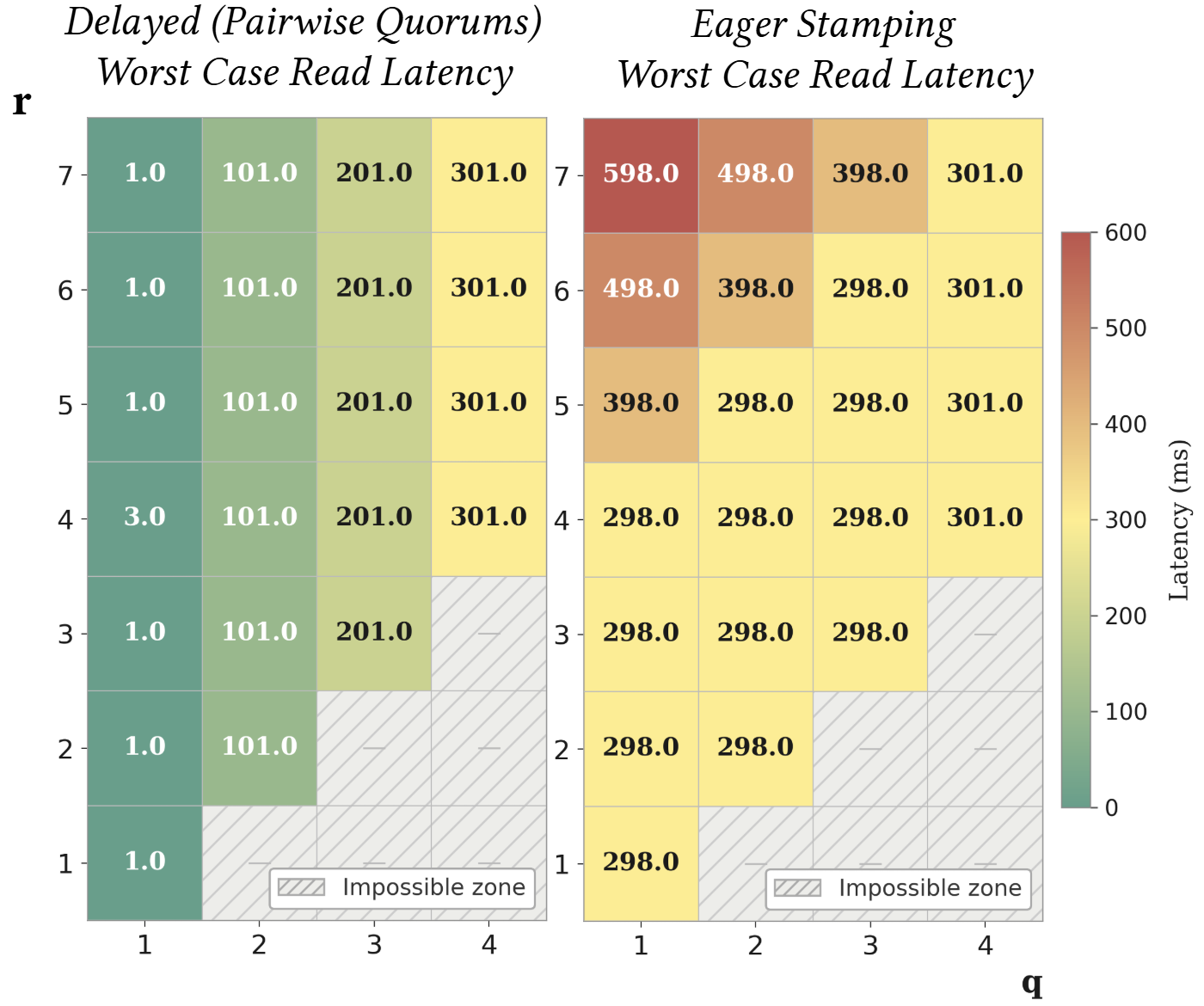}
    \caption{Measured worst case read latency (reads at leader $A$) over the $(r,q)$ space on the Base Network: Pairwise Quorums (delayed, left) vs. EdgePQR (eager, right), reproducing Figure \ref{fig:n_del_full_heatmap} and confirming delayed stamping's advantage.}
    \label{fig:n_eval_pwq_repeat}
\end{figure}

Finally, Figure \ref{fig:n_eval_pwq_repeat} measures worst case read latency across the $(r,q)$ space on the Base Network, reading at leader $A$. The left heatmap uses Pairwise Quorums and the right EdgePQR, both over $5000$ operations at a $5ms$ interval. To drive the worst case, we alternate reads and writes at the leader, so every read overlaps a freshly initiated, still-in-progress write. The measurements again reproduce Figure \ref{fig:n_del_full_heatmap} to within a few milliseconds per cell.

Together, these three experimental heatmaps validate our theoretical analyses and derived insights from sections \ref{subsec:categorization-analysis} and \ref{subsec:extending-analysis}, confirming the two core tradeoff axes in the configuration space, and the advantage of delayed stamping.

\subsection{Removing the Stamping-Variance Tradeoff}

\begin{figure}[htbp]
    \centering
    \includegraphics[width=0.4\textwidth]{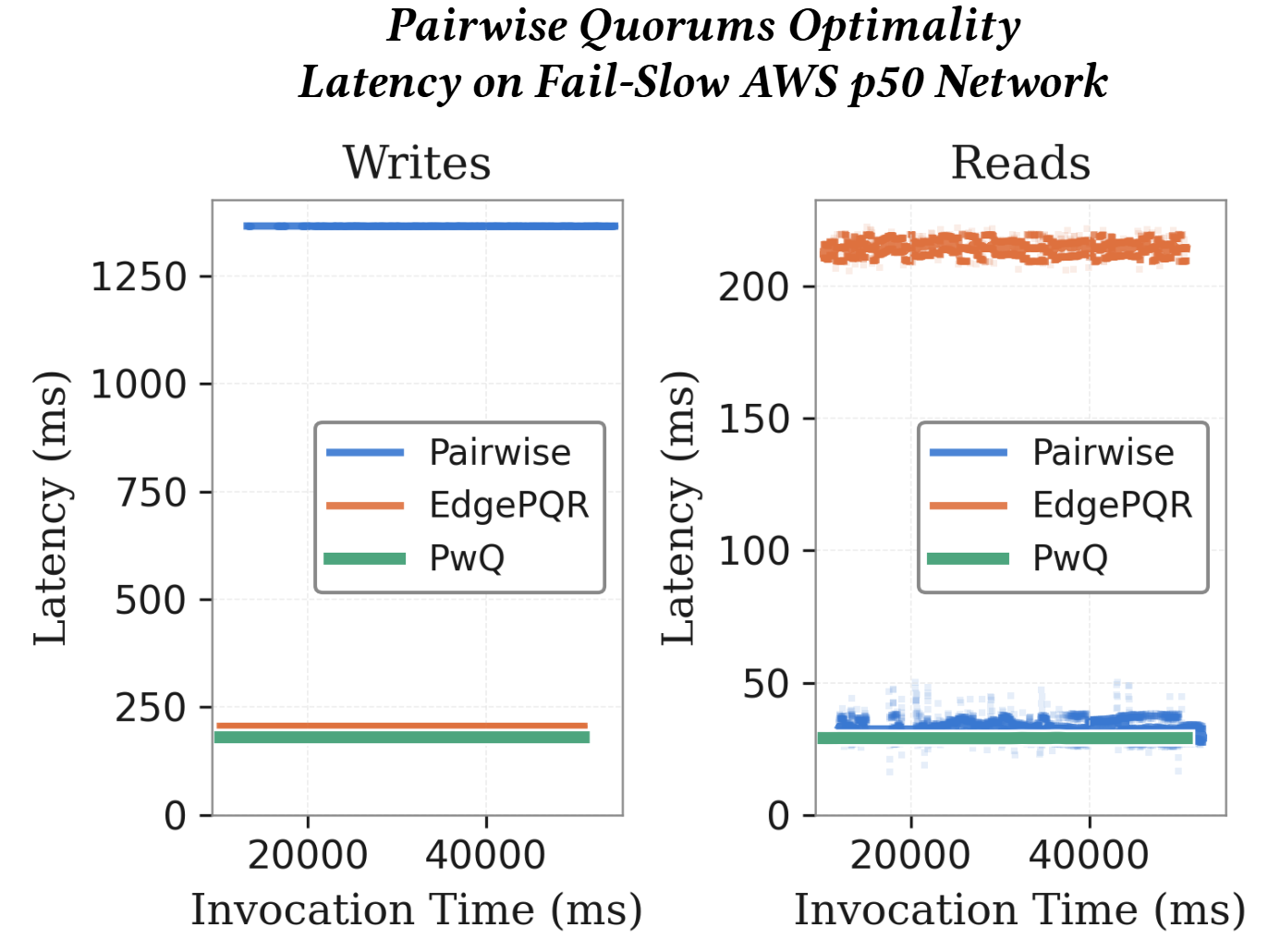}
    \caption{
    Per-operation latency over time on the AWS p50 network with one fail-slow reader. Pairwise Quorums keeps reads and writes fast, Pairwise pays the spike on writes, EdgePQR pays eager blockage on reads.}
    \label{fig:n_eval_pwq_wins_scatter}
\end{figure}

This experiment shows that Pairwise Quorums escapes the tradeoff between read blockage time and write tolerance to network variance that the existing alternatives are forced to make. On the emulated 5-node AWS p50 network we run three executions, one each for Pairwise, EdgePQR, and Pairwise Quorums (PwQ). We configure EdgePQR and PwQ to every size-two quorum ($q=2, r=5$), and Pairwise to every process a local reader. In each execution, all processes but the leader run 100\% reads, while the leader alternates 50\% reads and 50\% writes to drive worst case read latency through maximal write contention. We run $5000$ operations at a $5ms$ interval and make the second-furthest process from the leader (by p50 latency) fail-slow by adding $1200ms$ to its incoming messages. Figure \ref{fig:n_eval_pwq_wins_scatter} plots per-operation latency over time, side by side for reads at the process closest to the leader and writes at the leader.

The results expose the tradeoff. Pairwise reads quickly, but since every write must reach all local readers, every write incurs the fail-slow spike. EdgePQR's larger quorums let writes route around the slow process, yet its eager stamping makes every contended read pay the full worst case blockage. Pairwise Quorums alone escapes both: its size-two quorums absorb the spike on writes while its delayed stamping keeps reads fast, reaching the optimal corner that neither alternative can.

\subsection{The Value of Reconfiguration}

The previous experiment assumes the right configuration is known in advance. When conditions are unknown or change over time, no static configuration suffices, and Ditto reconfigures to track the optimum. In both experiments below, Ditto is hard-coded to reconfigure at the right time, since policy design is left to future work (Section \ref{sec:conclusion}).

\begin{figure}[htbp]
    \centering
    \includegraphics[width=0.35\textwidth]{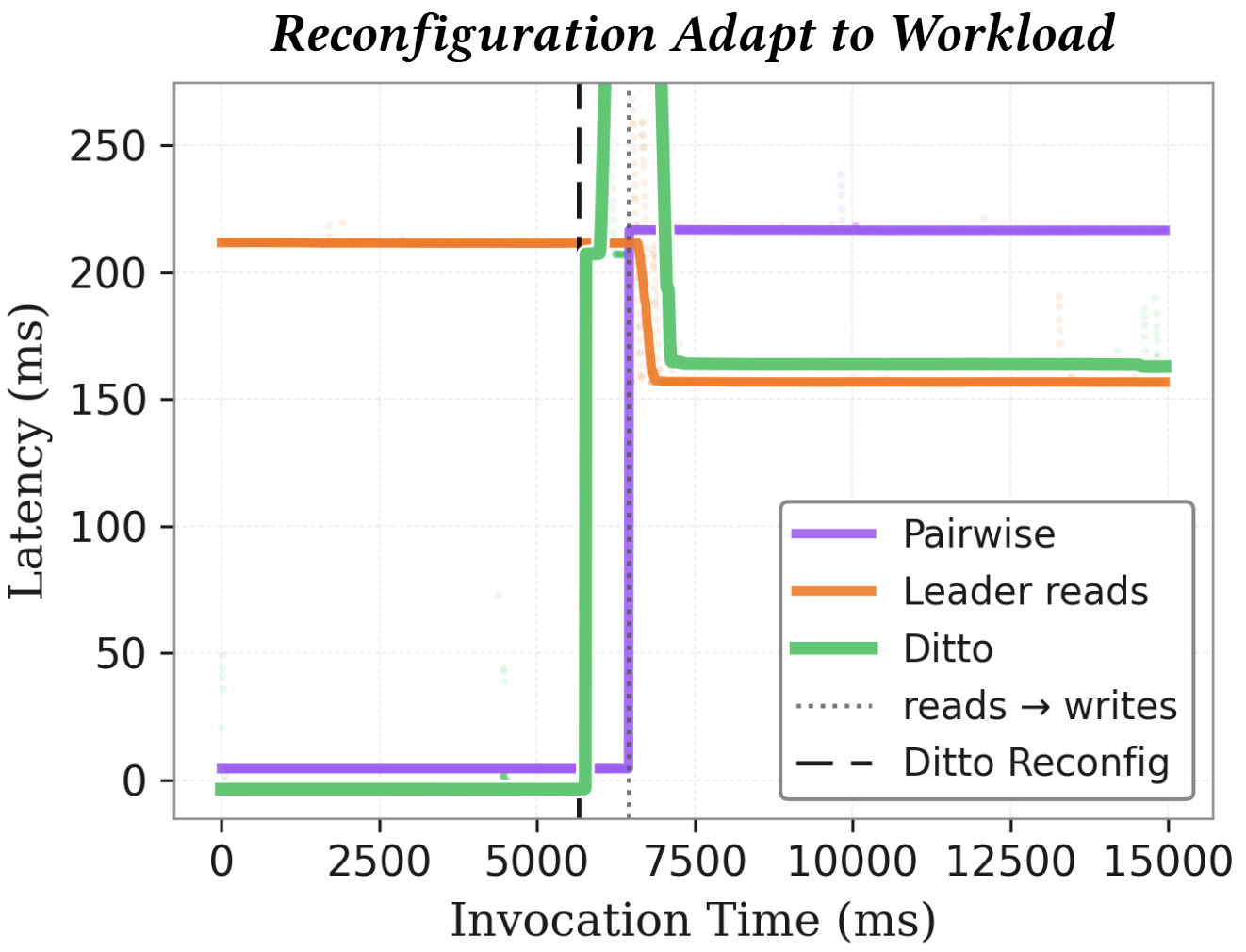}
    \caption{Operation latencies as the workload shifts from reads to writes on the AWS p50 network. Ditto reconfigures from the read-optimal to the write-optimal configuration, matching the best static choice in each phase.}
    \label{fig:n_eval_reconfig_workload}
\end{figure}

We first exercise reconfiguration along the read-write latency axis. On the emulated AWS p50 network we dispatch $3000$ reads at the process second-furthest from the leader, then $3000$ writes at the leader, each at a $5ms$ interval. We compare Ditto against Pairwise and Leader Reads, the read-optimal and write-optimal extremes. All three run delayed stamping, so any difference is due to quorum configuration alone. Pairwise and Leader Reads stay fixed, while Ditto starts read-optimal and reconfigures to Leader Reads, matching the better extreme in each phase. Figure \ref{fig:n_eval_reconfig_workload} plots operation latency over time; a dotted line marks the switch from all reads (the first 3000 operations) to all writes, and a dashed line marks where Ditto reconfigures. Apart from a brief spike as the new configuration installs -- an expected cost -- Ditto stays optimal throughout.

\begin{figure}[htbp]
    \centering
    \includegraphics[width=0.37\textwidth]{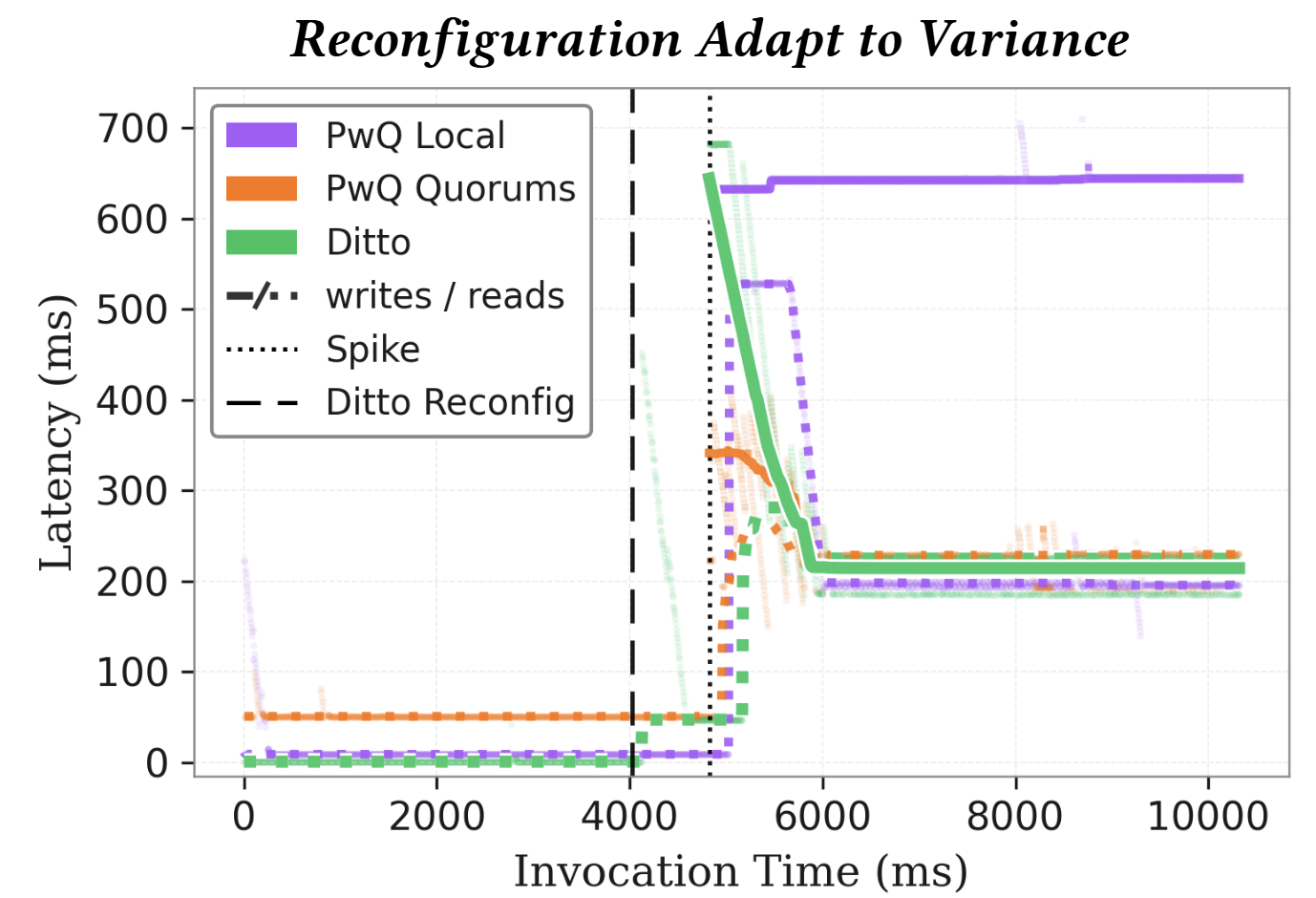}
    \caption{Per-operation latency as reads give way to concurrent reads and writes and a process turns fail-slow, on the AWS p50 network. Ditto reconfigures from read-optimal to variance-tolerant, matching the best static choice in each phase.}
    \label{fig:n_eval_reconfig_variance}
\end{figure}

We next exercise reconfiguration along the network-variance axis. On the emulated AWS p50 network we dispatch $3000$ reads at the process second-furthest from the leader (eu-central-1) at a $5ms$ interval, then spike the second-closest process (us-west-1) to fail-slow by adding $600ms$ to its incoming messages. After the spike, a further $3000$ reads at the second-furthest process run concurrently with $3000$ writes at the leader, both at a $5ms$ interval. Figure \ref{fig:n_eval_reconfig_variance} plots read latency (dotted lines) and write latency (solid lines) over time, with a vertical dotted line marking the spike and the end of the first $3000$ reads. We compare Ditto against two static Pairwise Quorums configurations: one with the four processes closest to the leader as local readers, and one with size-two quorums covering every process ($q=2, r=5$). The first is read-optimal, serving reads locally; the second tolerates any single fail-slow process, shielding writes from the variance at a modest cost to reads. Each static choice is stuck with its tradeoff, but Ditto reconfigures past it. Before the spike, both Ditto and local reads serve reads at $0ms$ while the quorum configuration sits at $50ms$; after the spike and reconfiguration settle, Ditto and the quorum configuration share a worst case read latency of $222ms$ and a write latency of $222ms$, while local reads holds marginally faster reads ($188ms$) but pays heavily on writes ($640ms$). Ditto again absorbs a few seconds of reconfiguration cost, but maintains optimality. One might ask why not simply reconfigure the fail-slow process out rather than going to quorums. That is only possible after the fact and demands a fresh reconfiguration every time a process slows, whereas a quorum configuration tolerates any single fail-slow process with no further reconfiguration at only a minor cost to reads.

\section{Related Work}
\label{sec:related}
 
\paragraph{State machine replication.}
Our work builds on leader-based SMR, including Paxos~\cite{lamport1998part,lamport2001paxos} and its multi-decree form MultiPaxos~\cite{van2015paxos}, Viewstamped Replication~\cite{oki1988viewstamped}, and Raft~\cite{ongaro2014search}. Fundamental limits on these protocols come from impossibility results such as FLP~\cite{fischer1985impossibility} and CAP~\cite{gilbert2002brewer}, and, for read-only transactions, SNOW~\cite{lu2016snow}. Leaderless and multi-leader protocols such as Mencius~\cite{mao2008mencius}, EPaxos~\cite{moraru2013there,tollman2021epaxos}, and SwiftPaxos~\cite{ryabinin2024swiftpaxos} rotate or remove the leader to improve wide-area latency and load balance, while Clock-RSM~\cite{du2014clock} and Nezha~\cite{geng2022nezha} order commands using synchronized clocks. We assume a stable leader and do not address these protocols.
 
\paragraph{Linearizable reads.}
Section~\ref{sec:background} places the leader-based read algorithms~\cite{chandra2007paxos,moraru2014paxos,chandra2016algorithm,bi2022parameterized,thiessen2025asymmetric,charapko2019linearizable,guarnieri2023linearizable} in our framework. Raft offers analogous leader reads, either confirming leadership with a round of heartbeats (ReadIndex) or relying on a leader lease~\cite{ongaro2014consensus}. Several systems outside leader-based SMR also serve linearizable reads away from the leader. Megastore~\cite{baker2011megastore} serves local reads at any replica whose coordinator vouches that it is up to date, and a write must invalidate the coordinators of replicas that miss it. Hermes~\cite{katsarakis2020hermes} lets every replica read locally by broadcasting invalidations on each write, and relies on an external membership service for fault tolerance. Chain replication~\cite{van2004chain} serves reads at the tail, and CRAQ~\cite{terrace2009object} lets any node serve reads, querying the tail only for objects with an uncommitted version. Gryff~\cite{burke2020gryff} pairs consensus for read-modify-writes with shared-register quorum reads. In our terms, Megastore and Hermes resemble invalidation over local read quorums, and CRAQ trades read blockage for a round trip to the tail. Because these systems order writes per object or depend on external coordination, they fall outside our model and we do not evaluate them. Bodega \cite{hu2026bodega} orders writes in a single log but serves local reads at per-key responders protected by all-to-all roster leases, and a read waits only on in-flight writes to the same key. This per-key conflict detection relies on key-value semantics that our model, a single replicated object updated by read-modify-writes, does not capture, so Bodega also falls outside it.
 
\paragraph{Quorum systems.}
Quorum systems have a long history in replicated data~\cite{naor1998load,cheung1992grid}, including dynamic quorum adjustment~\cite{herlihy1987dynamic}. Flexible Paxos~\cite{howard2016flexible} showed that consensus only needs leader-election quorums to intersect replication quorums, which WPaxos~\cite{ailijiang2020wpaxos} and DPaxos~\cite{nawab2018dpaxos} exploit to place quorums near users; DPaxos can also move its leader-election zones at runtime. Quoracle~\cite{whittaker2021read} and Pando~\cite{uluyol2020near} select a read-write quorum system before deployment by optimizing load, latency, or cost. Our analysis plays a similar role for the latency and variance tolerance of linearizable reads, including the effect of the stamping mechanism, and Ditto acts on it at runtime. An offline optimizer such as Quoracle could serve as one of Ditto's reconfiguration policies.
 
\paragraph{Reconfiguration.}
Reconfiguration in consensus has been studied extensively, including Raft's joint consensus~\cite{ongaro2014search}, Vertical Paxos~\cite{lamport2009vertical}, SMART~\cite{lorch2006smart}, and general methods for reconfiguring state machines~\cite{lamport2010reconfiguring}. RAMBO~\cite{lynch2002rambo} and DynaStore~\cite{aguilera2011dynamic} reconfigure the replica sets and quorums of atomic read/write storage, with RAMBO using the old and new configurations together during a transition. These works change the membership or quorums of the replication protocol itself, whereas Ditto reconfigures only the read quorum system on top of an unchanged SMR instance.
 
\paragraph{Fail-slow faults and synchronized clocks.}
Fail-slow and gray failures are common in production systems~\cite{gunawi2018fail,huang2017gray}. Copilot replication~\cite{ngo2020tolerating} tolerates one slow replica by having two distinguished replicas, a pilot and a copilot, both order every command, so either can make progress when the other slows. Ditto instead tolerates slow processes by choosing read quorums that let writes route around them. Synchronized clocks offer another path to fast reads. Spanner~\cite{corbett2013spanner} waits out clock uncertainty at commit so that sufficiently up-to-date replicas can serve consistent reads, and CockroachDB's global tables~\cite{vanbenschoten2022enabling} schedule writes at future timestamps so that any replica can serve strongly consistent reads locally, an idea close to the promise time of delayed stamping. Both rely on bounded clock skew, whereas delayed stamping needs only bounded drift; exploiting bounded skew, as BHT~\cite{bi2022parameterized} does, is future work.

\section{Conclusion}
\label{sec:conclusion}

We presented a unified framework for reasoning about linearizable reads algorithms, decomposing each into a choice of read-write quorum system and a stamping mechanism and placing all existing algorithms in a single space. A mathematical analysis of write and read latency over this space showed that no configuration dominates: the quorum system sets two tradeoffs, read versus write latency and read versus write tolerance to network variance. Meanwhile the stamping mechanism has a single best choice, delayed stamping, that prior work could apply only to local reads. We closed that gap with \textit{Pairwise Quorums}, which extends delayed stamping to arbitrary quorum configurations, and built \textit{Ditto} on top of it to reconfigure $Q^R$ at runtime. Under our model, Ditto subsumes every existing algorithm, so a deployment now needs to consider only Ditto and can use our analysis to guide its reconfiguration.

Several directions remain open. First, GPS clocks \cite{corbett2013spanner} make bounded clock skew, rather than only bounded drift, a practical assumption. Bounded skew keeps clocks synchronized within a fixed bound regardless of distance, which under delayed stamping could give near-zero blockage at every process, not just those near the leader, a property BHT \cite{bi2022parameterized} already claims. Extending our analysis to a GPS-clock stamping mechanism is a natural next step. Second, our analysis can guide reconfiguration but says nothing about \textit{when} to reconfigure, or how to balance conservative against aggressive strategies. Third, Ditto assumes reconfigurations are infrequent and does not optimize their cost; edge deployments, with frequent failures and a volatile network, may demand far faster reconfiguration.

A natural next step is also read throughput, which our analysis does not address. It depends on the read contact strategy, on message sizes that in Pairwise Quorums grow with in-progress writes, and on queueing effects, so it calls for a different treatment than latency. The contact strategy alone carries a further tradeoff: the \textit{full broadcast} we assume always finds a process's closest quorum under variance but floods the network on every read, whereas the \textit{thrifty} optimization \cite{tollman2021epaxos} contacts only the known-closest quorum, lowering load at the cost of slower adaptation to variance. Characterizing read throughput across the space, and tuning along it, is a promising direction.

\balance
\bibliographystyle{ACM-Reference-Format}
\bibliography{main}

\end{document}